\documentclass[final]{JHEP3} 

\JHEPspecialurl{http://jhep.sissa.it/JOURNAL/JHEP3.tar.gz}

\usepackage{epsfig,multicol,bbm,amsmath,graphicx}

\newcommand\fverb{\setbox\fverbbox=\hbox\bgroup\verb}
\newcommand\fverbdo{\egroup\medskip\noindent%
			\fbox{\unhbox\fverbbox}\ }
\newcommand\fverbit{\egroup\item[\fbox{\unhbox\fverbbox}]}
\newbox\fverbbox

\def\chioi{\tilde{\chi}^0_1}
\def\chioii{\tilde{\chi}^0_2}

\def\slr{\tilde{l}_R}
\def\sql{\tilde{q}_L}

\def\ETM{E_T^{\rm miss}}

\def\mll{m_{ll}}
\def\mllmax{m_{ll}^{\rm max}}
\def\mqln{m_{ql_n}}
\def\mqlnmax{m_{ql_n}^{\rm max}}
\def\mqlf{m_{ql_f}}
\def\mqlfmax{m_{ql_f}^{\rm max}}
\def\mqlfmaxmllzero{m_{ql_f}^{{\rm max}(0)}}
\def\mqlfmaxmll{m_{ql_f}^{{\rm max}(\mll)}}
\def\mqlfmaxmqln{m_{ql_f}^{{\rm max}(\mqln)}}
\def\mqll{m_{qll}}
\def\mqllhi{m_{qll{\rm (hi)}}}
\def\mqllmax{m_{qll}^{\rm max}}
\def\mqllmaxmll{m_{qll}^{{\rm max}(\mll)}}
\def\mqllmin{m_{qll}^{\rm min}}
\def\mqlhi{m_{ql{\rm (hi)}}}
\def\mqlhimax{m_{ql{\rm (hi)}}^{\rm max}}
\def\mqlhimaxmll{m_{ql{\rm (hi)}}^{{\rm max}(\mll)}}
\def\mqllo{m_{ql{\rm (lo)}}}
\def\mqllomax{m_{ql{\rm (lo)}}^{\rm max}}
\def\mqllomaxmll{m_{ql{\rm (lo)}}^{{\rm max}(\mll)}}
\def\mqleq{m_{ql{\rm (eq)}}}
\def\mqleqmax{m_{ql{\rm (eq)}}^{\rm max}}

\def\mij{m_{ij}}

\def\mjk{m_{jk}}
\def\mki{m_{ki}}
\def\mXqmax{m_{Xq}^{\rm max}}
\def\mql{m_{ql}}

\def\msql{m_{\sql}}
\def\mchioii{m_{\chioii}}
\def\mslr{m_{\slr}}
\def\mchioi{m_{\chioi}}

\def\max{{\rm max}}

\def\MT2{M_{T2}}

\def\m0{${0}$}

\title{Supersymmetric particle mass measurement with invariant mass correlations}

\author{Davide Costanzo\footnote{{\tt davide.costanzo@cern.ch}} \mbox{} and Daniel R. Tovey\footnote{{\tt daniel.tovey@cern.ch}}\\ Department of Physics and Astronomy, \\
	University of Sheffield, Hounsfield Road, \\ Sheffield S3 7RH,
	UK}

\received{} 		
\accepted{}		

\preprint{}

\abstract{The kinematic end-point technique for measuring the masses
of supersymmetric particles in R-Parity conserving models at hadron
colliders is re-examined with a focus on exploiting additional
constraints arising from correlations in invariant mass
observables. The use of such correlations is shown to potentially
resolve the ambiguity in the interpretation of quark+lepton end-points
and enable discrimination between sequential two-body and three-body
lepton-producing decays. The use of these techniques is shown to
improve the SUSY particle mass measurement precision for the SPS1a
benchmark model by at least 20-30\% compared to the conventional
end-point technique.}


\begin{document} 


\section{Introduction}\label{sec1}
Measurement of SUSY particle (`sparticle') masses in R-parity
conserving SUSY events at hadron colliders such as the LHC is
complicated by the pair production of invisible Lightest
Supersymmetric Particles (LSPs), one of which terminates each SUSY
particle decay chain. The presence of these LSPs prevents direct
measurement of the masses of the sparticles participating in the decay
chains via peaks in the invariant mass distributions of the observable
SUSY decay products.

Several approaches to resolving this problem have been
documented. These approaches may be grouped into three general
categories. The first category consists of techniques which use the
measurement of kinematic end-points in distributions of the invariant
masses of combinations of visible SUSY decay products (jets, leptons
etc.) from a single decay chain in each event. Given a long decay
chain sufficient constraints may be obtained to solve for the
individual masses
\cite{Hinchliffe:1996iu,Allanach:2000kt,Gjelsten:2004ki,Miller:2005zp}. The
second category consists of techniques which use the measurement of
kinematic end-points in distributions of observables constructed from
the transverse momenta of SUSY decay products from two identical decay
chains appearing in each event
\cite{Lester:1999tx,Barr:2003rg,Cho:2007qv,Tovey:2008ui}. These
techniques can provide constraints on combinations of sparticle masses
from very short (one-step) chains and can in principle lead to
constraints on individual masses
\cite{Cho:2007qv,Tovey:2008ui,Gripaios:2007is,Barr:2007hy,Cho:2007dh}. The
third category consists of hybrid methods which seek to use optimally
both transverse momentum and invariant mass constraints from a subset
of events in which the same long decay chain appears in both `legs' of
each event (e.g. \cite{Cheng:2007xv,Nojiri:2007pq}). Such techniques
can offer an improvement on the mass constraints obtained by other
techniques, but may require significant integrated luminosity to
generate a sufficiently large sample of the specific events of
interest.

This paper shall focus on the first category of mass measurement
techniques -- that involving the use of kinematic end-point
constraints. The goal of the paper is to re-examine the constraints
which may be obtained from kinematic end-points in the three-step
sequential two-body decay chains most usually studied, with a focus on
correlations between invariant mass observables. This will lead to a
simplification of the technique and an improvement in the possible
mass measurement precision. Similar techniques have been studied
recently for the different case of events in which SUSY higgs
particles decay symmetrically via two two-step sequential two-body
decays \cite{Huang:2008ae,Bisset:2008hm}. Correlations between
invariant masses were also exploited in Ref.~\cite{Cheng:2007xv}.

Section 2 commences with a review of the conventional end-point
technique, while Section 3 re-interprets the technique from a
geometric perspective. Section 4 examines the additional constraints
which can be obtained by making use of correlations between invariant
mass observables while Section 5 examines how these constraints may be
used in principle to determine individual sparticle masses. Section 6
presents a case study in which the new techniques are used to measure
sparticle masses for the SPS1a benchmark model with fast detector
simulation. Section 7 concludes and outlines some possible directions
for future work.

\section{Kinematic constraints from three-step sequential two-body decay chains}
\label{sec2}

\subsection{Definition of decay chain}
\label{sec2.1}
In many regions of SUSY parameter space the sparticle mass hierarchy
is such that heavy strongly interacting sparticles produced in the
initial proton-proton interaction can decay via a 3-step sequential
two-body decay chain of the form:
\begin{equation}
\label{eqn1}
  \delta \rightarrow \gamma  c \rightarrow \beta  bc \rightarrow \alpha abc,
\end{equation}
where Greek letters denote SUSY states, Roman letters denote visible
SM decay products and $\alpha$ is the LSP. The canonical example of
such a decay chain is the decay
\begin{equation}
\label{eqn2}
  \sql \rightarrow \chioii  q \rightarrow \slr l_nq \rightarrow \chioi l_fl_nq,
\end{equation}
where a left-squark decays via a chain consisting of the
next-to-lightest neutralino, a right-slepton and a lightest neutralino
(LSP), in the process emitting a quark ($q$) and two opposite-sign
same-flavour leptons ($l_n$, $l_f$). The notation $l_n$($l_f$) is
conventionally used to denote the lepton emitted nearest to (furthest
from) the quark in the decay chain. In practice $l_n$ is
indistinguishable from $l_f$ which leads to ambiguities in the mass
constraints obtained from this technique.

In this paper we shall illustrate mass measurment techniques at both
parton and detector-level using the SPS1a benchmark SUSY model
\cite{Allanach:2002nj,Weiglein:2004hn}. The relevant mean sparticle
masses for this model are $\mchioi$ $=$ 96.05 GeV, $\mslr$ $=$ 142.97
GeV, $\mchioii$ $=$ 176.82 GeV and 491.9 GeV $<$ $\msql$ $<$ 543.0
GeV. In the $\sql$ case we assume dominant light squark production and
a common light squark mass of 540 GeV when constructing kinematic
bounds. The limited validity of this simplifying assumption, together
with the non-zero widths of the squarks (especially) causes some
parton-level distributions to `leak' beyond the expected bounds (see
e.g. Figure~\ref{fig2a} below). For this study {\tt HERWIG 6.4}
\cite{Corcella:2000bw,Moretti:2002eu} was used to generate 300k (100
fb$^{-1}$) parton-level SUSY events filtered to require at least two
final state leptons in each event.

\paragraph{A note on notation} In this paper we shall frequently refer
both to global bounds on invariant mass distributions (`global
end-points') and to bounds obtained from subsets of events satisfying
certain selection criteria imposed on other invariant mass quantities
(`conditional end-points'). In common with previous studies we shall
refer to the former with quantities with a single superscript `max' or
`min'. We shall refer to the latter with quantities with a superscript
`max($m$)' where $m$ represents the invariant mass upon which the
selection criteria are imposed.

\subsection{Invariant mass end-points}
\label{sec2.2}

The end-point technique starts by constraining combinations of masses
of sparticles appearing in the decay chain given by Eqn.~(\ref{eqn2})
by using kinematic bounds (`end-points') in the distributions of the
invariant masses of combinations of $q$, $l_n$ and $l_f$. Maxima in
the distributions of two-body invariant masses $\mij$ are obtained
when visible particles $i$ and $j$ are emitted in opposite directions
in the rest frame of the $\chioii$ sparticle in each chain. Because
each decay is two-body the momenta of $i$ and $j$ are fixed in the
rest frames of their respective parents. Consequently there is a
one-to-one mapping between the opening angles of the $i$ and $j$ in
the rest frame of the $\chioii$ and $\mij$. Setting this angle to its
maximum value of $\pi$ radians therefore allows the position of the
kinematic end-points to be determined analytically. The formulae are
(see e.g. Ref.~\cite{Hinchliffe:1996iu}):
\begin{eqnarray}
\big(\mllmax\big)^2 &=& \frac{\big(\mchioii^2-\mslr^2\big)\big(\mslr^2-\mchioi^2\big)}{\mslr^2}, \label{eqn3}\\
\big(\mqlnmax\big)^2 &=& \frac{\big(\msql^2-\mchioii^2\big)\big(\mchioii^2-\mslr^2\big)}{\mchioii^2}, \label{eqn4}\\
\big(\mqlfmax\big)^2 &=& \frac{\big(\msql^2-\mchioii^2\big)\big(\mslr^2-\mchioi^2\big)}{\mslr^2}. \label{eqn5}
\end{eqnarray} 
For the following discussion it will also be useful to define a
quantity $\mqlfmaxmllzero$ which measures the maximum value of $\mqlf$
when the two leptons are co-linear and hence $\mll=0$:
\begin{equation}
\label{eqn29}
\big(\mqlfmaxmllzero\big)^2 = \frac{\big(\msql^2-\mchioii^2\big)\big(\mslr^2-\mchioi^2\big)}{\mchioii^2}.
\end{equation}

The ambiguity of assigning observed lepton momenta to $l_n$ and $l_f$
conventionally prevents direct measurement of $\mqlnmax$ and
$\mqlfmax$. Instead, for each event the two possible lepton+quark(jet)
invariant masses are ordered to generate $\mqlhimax$ and $\mqllomax$
with analytical end-point positions at \cite{Allanach:2000kt}
(following Ref.~\cite{Gjelsten:2004ki}):
\begin{eqnarray}
\big(\mqllomax,\mqlhimax\big) &=& \left\{
\begin{array}{llcc}
\big(\mqlnmax,\mqlfmax\big) & 
\quad {\rm for } \quad
& 2\mslr^2 > \mchioi^2+\mchioii^2 > 2\mchioi\mchioii & \quad {\rm (A1)} \\[4mm]
\big(\mqleqmax,\mqlfmax\big) & 
\quad {\rm for } \quad
& \mchioi^2+\mchioii^2 > 2\mslr^2 > 2\mchioi\mchioii & \quad {\rm (A2)}\\[4mm]
\big(\mqleqmax,\mqlnmax\big) & 
\quad {\rm for } \quad
& \mchioi^2+\mchioii^2 > 2\mchioi\mchioii > 2\mslr^2 & \quad {\rm (A3)}
\end{array} 
\right\} \cr
&& \label{eqn6}
\end{eqnarray}
where 
\begin{equation}
\big(\mqleqmax\big)^2 =
\frac{\big(\msql^2-\mchioii^2\big)\big(\mslr^2-\mchioi^2\big)}{\big(2\mslr^2-\mchioi^2\big)}.
\label{eqn6a}
\end{equation}
The bound provided by $\mqleqmax$ arises from kinematic configurations
in which $\mql$ is maximised for $\mqln=\mqlf\equiv\mqleq$. Case A1
holds for the SPS1a model considered throughout this paper.

In addition to the two-body invariant mass end-points three-body
end-points may be observed using $\mqll$. The analytical formula for
the maximum value of the distribution of $\mqll$ is complicated by the
fact that it may be generated by any one of four kinematic
configurations, depending on the sparticle mass hierarchy. These
configurations are those in which two of $q$, $l_n$ and $l_f$ lie
co-linear in the $\chioii$ rest frame and one contra-linear (three
configurations) or their net momentum in the $\sql$ rest-frame is zero
(one configuration). Taking this ambiguity into account the maximum
value of the $\mqll$ distribution is found to lie at
\cite{Allanach:2000kt} (following Ref.~\cite{Gjelsten:2004ki}):
\begin{eqnarray}
\left(\mqllmax\right)^2 
&=& \left\{ 
\begin{array}{llcc}
\frac{\big(\msql^2-\mchioii^2\big)\big(\mchioii^2-\mchioi^2\big)}{\mchioii^2} 
& \ {\rm for } \quad 
& \frac{\msql}{\mchioii} > \frac{\mchioii}{\mslr}\frac{\mslr}{\mchioi} & \quad {\rm (B1)} \\[4mm]
\frac{\big(\msql^2\mslr^2-\mchioii^2\mchioi^2\big)\big(\mchioii^2-\mslr^2\big)}{\mchioii^2\mslr^2} 
& \ {\rm for } \quad
&  \frac{\mchioii}{\mslr} > \frac{\mslr}{\mchioi}\frac{\msql}{\mchioii} & \quad {\rm (B2)}\\[4mm]
\frac{\big(\msql^2-\mslr^2\big)\big(\mslr^2-\mchioi^2\big)}{\mslr^2} 
& \ {\rm for } \quad
& \frac{\mslr}{\mchioi} > \frac{\msql}{\mchioii}\frac{\mchioii}{\mslr} & \quad {\rm (B3)} \\[4mm]
\big(\msql-\mchioi\big)^2 
& \ \lefteqn{\rm otherwise} && \quad {\rm (B4)} 
\end{array} 
\right\}  \label{eqn8}
\end{eqnarray}
The first three cases above (B1--B3) occur if one of the three
co-linear configurations provides the maximum $\mqll$ value, with
cases B1, B2 and B3 corresponding respectively to configurations with
$\big\{\mqlf=\mqlfmaxmllzero,\mll=0,\mqln=\mqlnmax\big\}$,
$\big\{\mqlf=0,\mll=\mllmax,\mqln=\mqlnmax\big\}$ and
$\big\{\mqlf=\mqlfmax,\mll=\mllmax,\mqln=0\big\}$. The fourth case
(B4) occurs if the zero momentum configuration is allowed
kinematically. Case B1 holds for the SPS1a model considered throughout
this paper.

The $\mqll$ distribution also possesses a non-zero minimum when
$m_{ll} > 0$. Conventionally a cut requiring $\mll>\mllmax/\sqrt{2}$
is applied, leading to a threshold in the $\mqll$ distribution at
$\mqllmin$ given by\footnote{The value of $\mqllmin$ is dependent on
the condition imposed upon $\mll$ and therefore should strictly be
identified in this paper as $\mqll^{{\rm min}(\mll)}$, in light of the
discussion in Section~\ref{sec2.1}. We choose to retain the notation
$\mqllmin$ for consistency with earlier work however.} :
\begin{multline}
\big(\mqllmin\big)^2 = \frac{1}{4\mslr^2\mchioii^2}\Big[2\mslr^2\big(\msql^2-\mchioii^2\big)\big(\mchioii^2-\mchioi^2\big) \big.\\ \big .+\big(\msql^2+\mchioii^2\big)\big(\mchioii^2-\mslr^2\big)\big(\mslr^2-\mchioi^2\big)\big . \\ \big .-\big(\msql^2-\mchioii^2\big)\sqrt{\big(\mchioii^2+\mslr^2\big)^2\big(\mslr^2+\mchioi^2\big)^2-16\mchioii^2\mslr^4\mchioi^2}\Big]. \label{eqn10}
\end{multline}

Together these five mass constraints may be solved numerically
\cite{Hinchliffe:1996iu,Allanach:2000kt} or analytically
\cite{Miller:2005zp} to determine the four individual sparticle masses
$m_{\sql}$, $m_{\chioii}$, $m_{\slr}$ and $m_{\chioi}$. In either case
the ambiguities in the interpretation of $\mqlhimax$, $\mqllomax$ and
$\mqllmax$ can lead to multiple solutions and hence ambiguity in
constraints on the underlying SUSY model. This is discussed
extensively in e.g. Ref.~\cite{Miller:2005zp}.

\section{Geometric interpretation}
\label{sec3} 

The conventional approach to the derivation of kinematic end-point
constraints illustrated above focuses on physical configurations of
particle momenta which maximise or minimise invariant masses. The
constraints are thus generated when the momentum vectors of visible
particles are correlated in a particular manner, for instance parallel
or anti-parallel. Let us now generalise this argument to reinterpret
these correlations in a geometric context. Our aim will be to use
these momentum correlations to identify correlations between the
invariant masses constructed from the momenta. For convenience we
shall work in the rest frame of the $\chioii$ in the following
discussion, however the calculated invariant masses are clearly equal
to those measured in the laboratory frame.

Our starting point is the observation that extremal values of $\mqll$
are obtained when the momenta of $q$, $l_n$ and $l_f$ lie in the same
plane in the rest frame of the $\chioii$. Furthermore the extremal
values of $\mll$, $\mqln$ and $\mqlf$ can trivially also be obtained
when this is the case. Neglecting the masses of $q$, $l_n$ and $l_f$
we can write the three invariant mass combinations as:
\begin{eqnarray}
\mll &=& 2p_{l_n}p_{l_f}\big(1-\cos\theta_{ll}\vphantom{\theta_{ql_f}}\big),\label{eqn11}\\
\mqln &=& 2p_{q}p_{l_n}\big(1-\cos\theta_{ql_n}\vphantom{\theta_{ql_f}}\big),\label{eqn12}\\
\mqlf &=& 2p_{q}p_{l_f}\big(1-\cos\theta_{ql_f}\big),\label{eqn13}
\end{eqnarray}
where $p_i$ denotes the magnitude of the three-momentum of particle $i$ and
$\theta_{ij}$ denotes the opening angle between the three-momenta of
particles $i$ and $j$. When $q$, $l_n$ and $l_f$ all lie in one plane
we have the further relation:
\begin{equation}
\label{eqn13a}
\theta_{ll} + \theta_{ql_n} + \theta_{ql_f} = 2\pi.
\end{equation}

Now in the rest frame of the $\chioii$ the momentum magnitude $p_{l_n}$ is
determined from simple two-body kinematics to be the fixed quantity:
\begin{equation}
\label{eqn19}
p_{l_n}=\frac{\mchioii^2-\mslr^2}{2\mchioii}.
\end{equation}
The momentum magnitude $p_{q}$ is also fixed in this frame to be:
\begin{equation}
\label{eqn20}
p_{q}=\frac{\msql^2-\mchioii^2}{2\mchioii}.
\end{equation}
The magnitude of the momentum of $l_f$, $p_{l_f}$, is however not a constant in
this frame -- it is related to the fixed momentum magnitude $p_{l_f}'$
measured in the rest frame of the $\slr$ by a Lorentz transformation
dependent on $\theta_{ll}$:
\begin{eqnarray}
\label{eqn21}
p_{l_f}' & \equiv& \frac{\mslr^2-\mchioi^2}{2\mslr} \cr
         & = & \gamma p_{l_f}\big(1+\beta\cos\theta_{ll}\big),
\end{eqnarray}
where the boost factor $\beta$ is given by:
\begin{equation}
\label{eqn21a}
\beta = \frac{\mchioii^2-\mslr^2}{\mchioii^2+\mslr^2},
\end{equation}
and $\gamma=(1-\beta^2)^{-1/2}$.

It is convenient now to define three dimensionless mass coordinates $x$, $y$ and $z$:
\begin{eqnarray}
x=\frac{\mqln}{\sqrt{p_{q}p_{l_n}}},\label{eqn14}\\
y=\frac{\mqlf}{\sqrt{p_{q}p_{l_f}'}},\label{eqn15}\\
z=\frac{\mll}{\sqrt{p_{l_n}p_{l_f}'}}, \label{eqn16}
\end{eqnarray}
together with a dimensionless mass ratio $r\equiv \mslr/\mchioii$.
We can now use Eqns.~(\ref{eqn11})--(\ref{eqn13}) to express
Eqn.~(\ref{eqn13a}) purely in terms of invariant masses $\mij$ and
momenta and then substitute from Eqns.~(\ref{eqn14})--(\ref{eqn16}) to
obtain a single relation between $x$, $y$ and $z$:
\begin{equation}
\label{eqn26}
\left(x^2\left(r+\frac{1}{4}\left(1-r^2\right)z^2\right)-y^2-z^2\right)^2+(x^2-4)y^2z^2=0.
\end{equation}

Eqn.~(\ref{eqn26}) defines a surface $\Sigma_{qll}$ in the
3-dimensional $x-y-z$ space. This surface bounds the region within
which all decay chains of the type Eqn.~(\ref{eqn2}) must lie: events
in which the visible decay products are coplanar lie on the surface,
while acoplanar events lie within the volume. The surface is plotted
in Figure~\ref{fig1} for the SPS1a benchmark model. A key feature to
note is that $\Sigma_{qll}$ intersects the planes with $x$, $y$ or $z$
zero along lines, since when any one $\mij$ value is zero, particles
$i$ and $j$ must be co-linear and hence $\mjk$ must be uniquely
determined by $\mki$ (and vice versa). Note in addition that if
$\Sigma_{qll}$ were plotted in $\mqln-\mqlf-\mll$ space rather than in
$x-y-z$ space the radial distance from the origin of a point on
$\Sigma_{qll}$ would give $\mqll$, and hence in these coordinates
$\mqllmax$ represents the greatest radial distance of a point on
$\Sigma_{qll}$ from the origin. Eqn.~(\ref{eqn26}) shall prove useful
for deriving some of the kinematic bounds in two- and three-dimensions
discussed below however it is potentially also useful in its own right
-- for instance for selecting events with the decay chain
Eqn.~\ref{eqn2} and associating decay products with specific steps in
that decay chain once the masses of the sparticles are known.

\FIGURE[ht]{
\epsfig{file=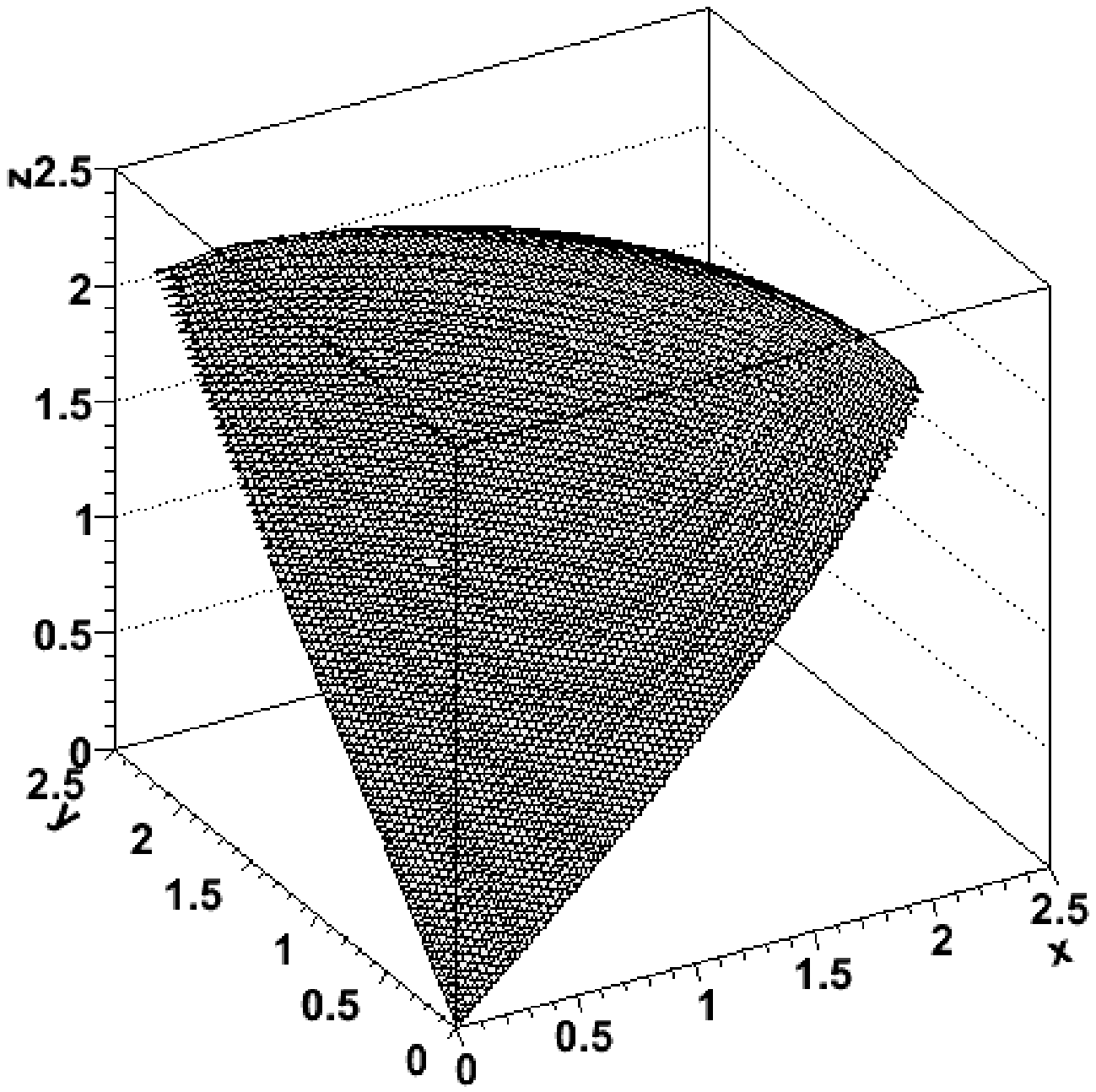,height=2.5in}
\epsfig{file=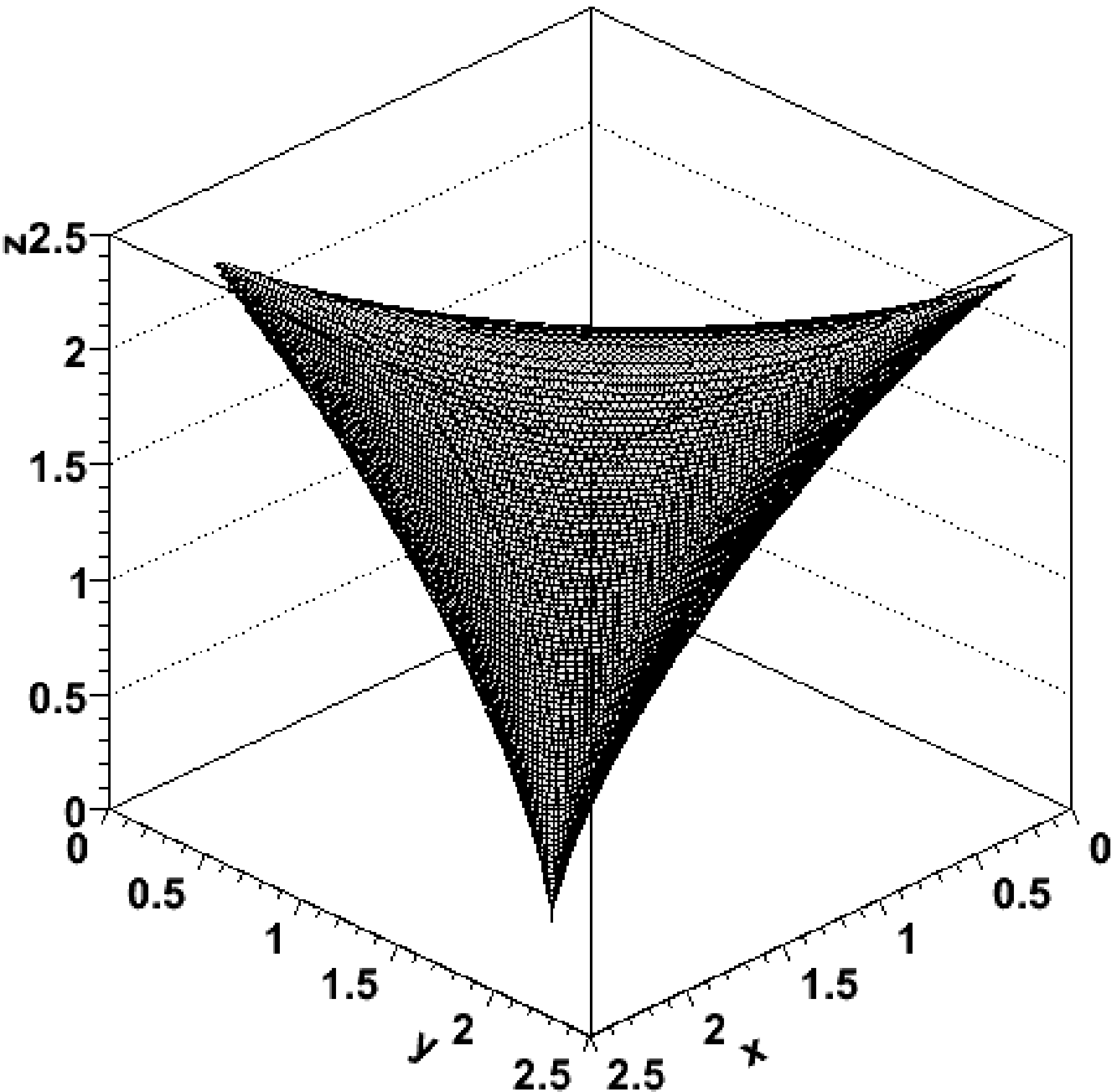,height=2.5in}
\caption{\label{fig1} $\Sigma_{qll}$ surface for the SPS1a benchmark
SUSY model plotted in $x-y-z$ space and viewed
from two different perspectives.} }

\section{Additional kinematic constraints}
\label{sec4}

\subsection{Global end-point constraints}
\label{sec4.1}
Constraints on the masses of the sparticles participating in the decay
chain listed in Eqn.~(\ref{eqn2}) can be obtained by measuring the
surface $\Sigma_{qll}$ defined by Eqn.~(\ref{eqn26}). The techniques
described in Section~\ref{sec2.2} rely on integrating out all but one
invariant mass observable and then measuring the global end-point in the
resulting mass distribution. It should be clear from the
discussion of Section~\ref{sec3} however that there is potentially
much more information contained in Eqn.~(\ref{eqn26}) than can be
exploited with this relatively simple technique. In particular the
shape of this surface can be measured in more detail by exploiting
correlations between invariant mass observables. Such correlations are
partially exploited by the definition of the $\mqllmin$ end-point in
Eqn.~(\ref{eqn10}), in which the integration takes place only over a
limited range in $\mll$, however the possibilities presented by this
technique are more numerous, as shall be illustrated in the following
discussion.

\subsection{Conditional end-point constraints with two-dimensional correlations}
\label{sec4.2}
The first class of additional constraints can be obtained by
integrating over only one, rather than two, of the three
degrees-of-freedom used to define the surface $\Sigma_{qll}$ in
Eqn.~(\ref{eqn26}). The most familiar example of such a technique is
that used to obtain the $\mqllmin$ end-point discussed above. The
correlation between $\mqll$ and $\mll$ can be observed by plotting
$\mqll^2-\mll^2$ against $\mll^2$, shown at parton-level in
Figure~\ref{fig2a} for the SPS1a benchmark model. The vertical
kinematic bound in this figure is provided by $\mllmax$ (right-hand
vertical line), while the minimum of the one-dimensional distribution
obtained by integrating along the $\mll^2$ axis to the right of the
left-hand vertical line measures $(\mqllmin)^2-(\mllmax)^2/2$.
\FIGURE[ht]{ \epsfig{file=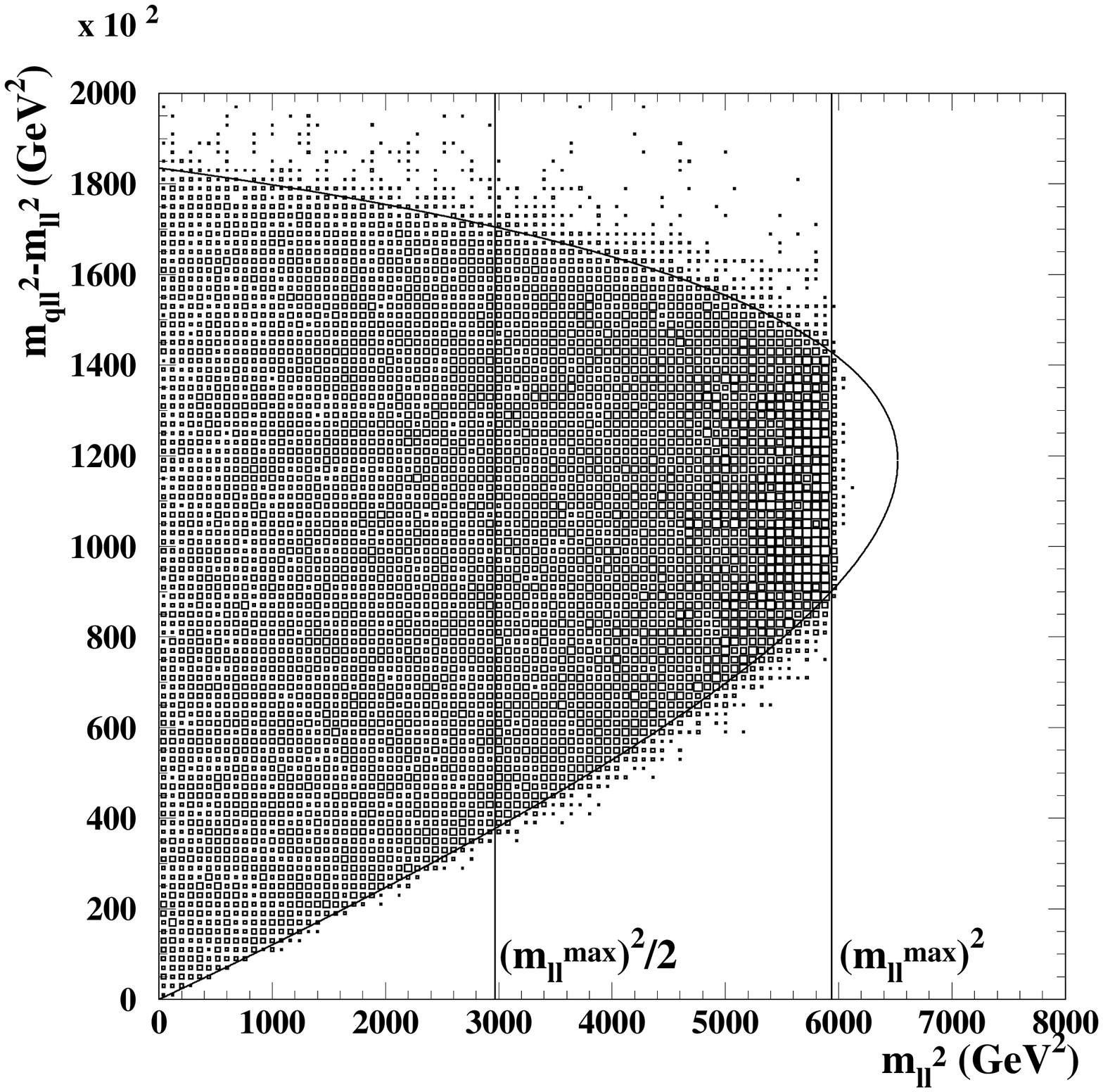,height=3.5in}
\caption{\label{fig2a} Parton-level two-dimensional invariant
mass-squared distribution for decay chain Eqn.~(\ref{eqn2}) for the
SPS1a benchmark model showing $\mqll^2-\mll^2$ plotted against
$\mll^2$. See text for explanation of bounds.} }

The curved bounds in Figure~\ref{fig2a} may be obtained from the
general formula (see e.g. Ref.~\cite{Lester:2001zx}) for the end-points in
the invariant mass distribution of two heavy visible decay products
$a$ and $b$ of a two-step sequential two-body decay chain $\gamma
\rightarrow \beta b \rightarrow \alpha ab$:
\begin{multline}
\label{eqn30}
\big(m_{ab}^{{\rm bound}(m_a,m_b)}\big)^2 = m_a^2 + m_b^2 + \frac{1}{2m_{\beta}^2}\Big[\big(m_{\beta}^2-m_{\alpha}^2+m_a^2\big)\big(m_{\gamma}^2-m_{\beta}^2-m_b^2\big) \vphantom{\sqrt{\lambda(m_{\beta})}} \big . \\ \big . \pm \sqrt{\lambda(m_{\beta},m_{\alpha},m_a)\lambda(m_{\gamma},m_{\beta},m_b)}\Big],
\end{multline}
where
\begin{equation}
\label{eqn31}
\lambda(m_i,m_j,m_k)  \equiv  \big(m_i^2-m_j^2+m_k^2\big)^2-4m_i^2m_k^2.
\end{equation}
When $a\equiv ll$ and $b\equiv q$ and $m_b=0$ this equation may be simplified to give:
\begin{equation}
\label{eqn32}
\big(\mqll^{{\rm bound}(\mll)}\big)^2=\mll^2 + \frac{\big(\msql^2-\mchioii^2\big)}{2\mchioii^2}\Big(\mchioii^2-\mchioi^2+\mll^2 \pm \sqrt{\lambda(\mchioii,\mchioi,\mll)}\Big),
\end{equation}
where the positive root corresponds to the usual expression for the
`$\mXqmax$' end-point with $X\equiv ll$ \cite{Hinchliffe:1996iu}. The two
roots of this equation correspond to the upper and lower curved bounds
in Figure~\ref{fig2a}. Setting $m_a=0$ with $a\equiv l_f$ and $b
\equiv ql_n$ in Eqn.~(\ref{eqn30}) also allows the equivalent bounds
in the $\mqll^2-\mqln^2$ versus $\mqln^2$ distribution to be obtained
(not shown). 

Plots similar to Figure~\ref{fig2a} can also be constructed from the
distributions of $\mqll^2-\mqln^2$ as a function of $\mqln^2$ and
$\mqll^2-\mqlf^2$ as a function of $\mqlf^2$. In the latter case
however the kinematic bounds are particularly difficult to describe
analytically. An alternative means of illustrating two-dimensional
correlations in which the kinematic bounds may be described more
easily is provided by the distributions of $\mij^2$ as functions of
$\mjk^2$ where $i$, $j$ and $k$ are drawn from $q$, $l_n$ and
$l_f$. The three possible distributions are shown in Figure~\ref{fig2}
at parton-level for the SPS1a model. Squared masses are plotted in
this and subsequent figures in this section (as in Figure~\ref{fig2a})
as this generates kinematic bounds which take on particularly simple forms.
\FIGURE[ht]{
\epsfig{file=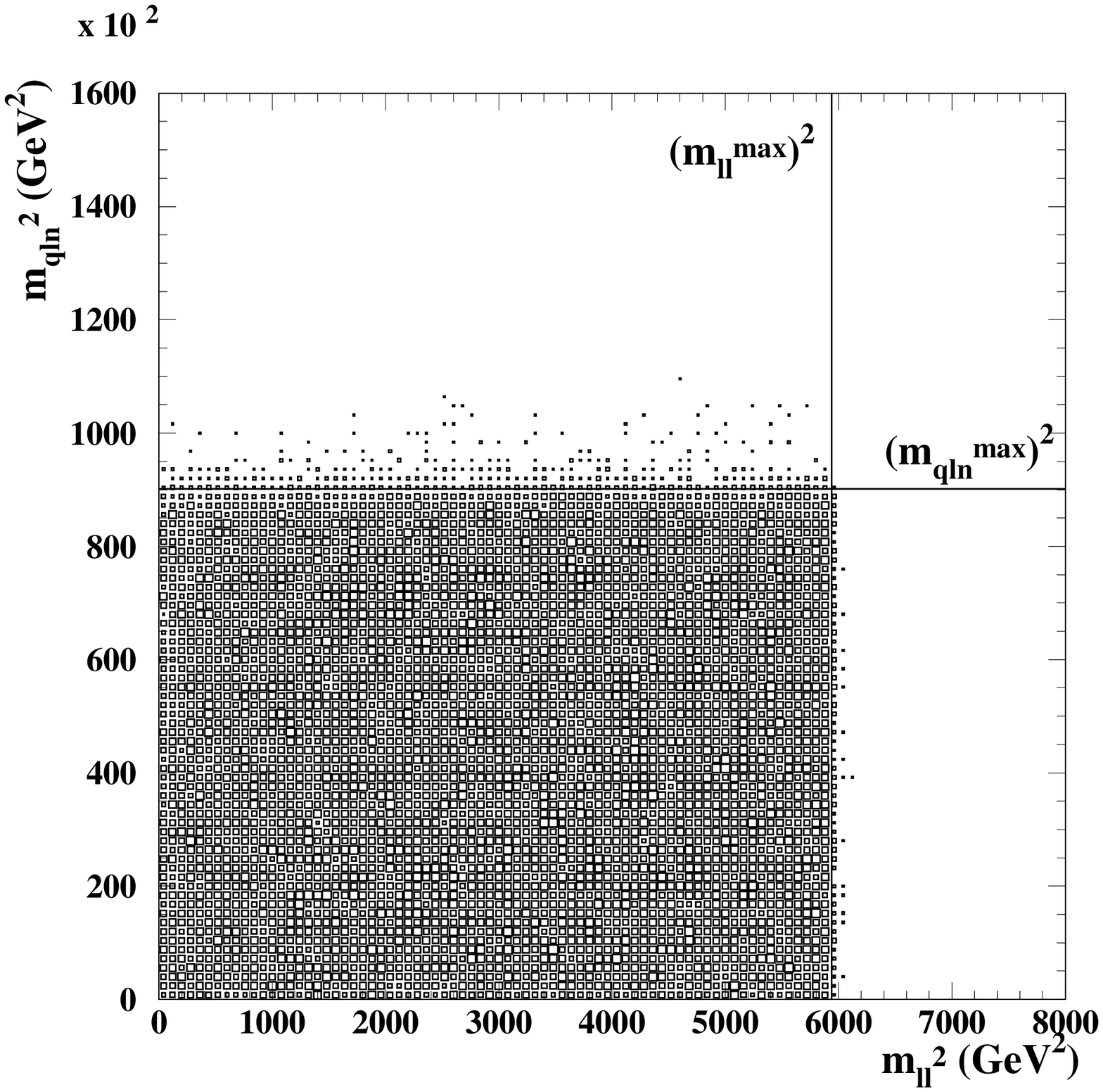,height=2.5in}
\epsfig{file=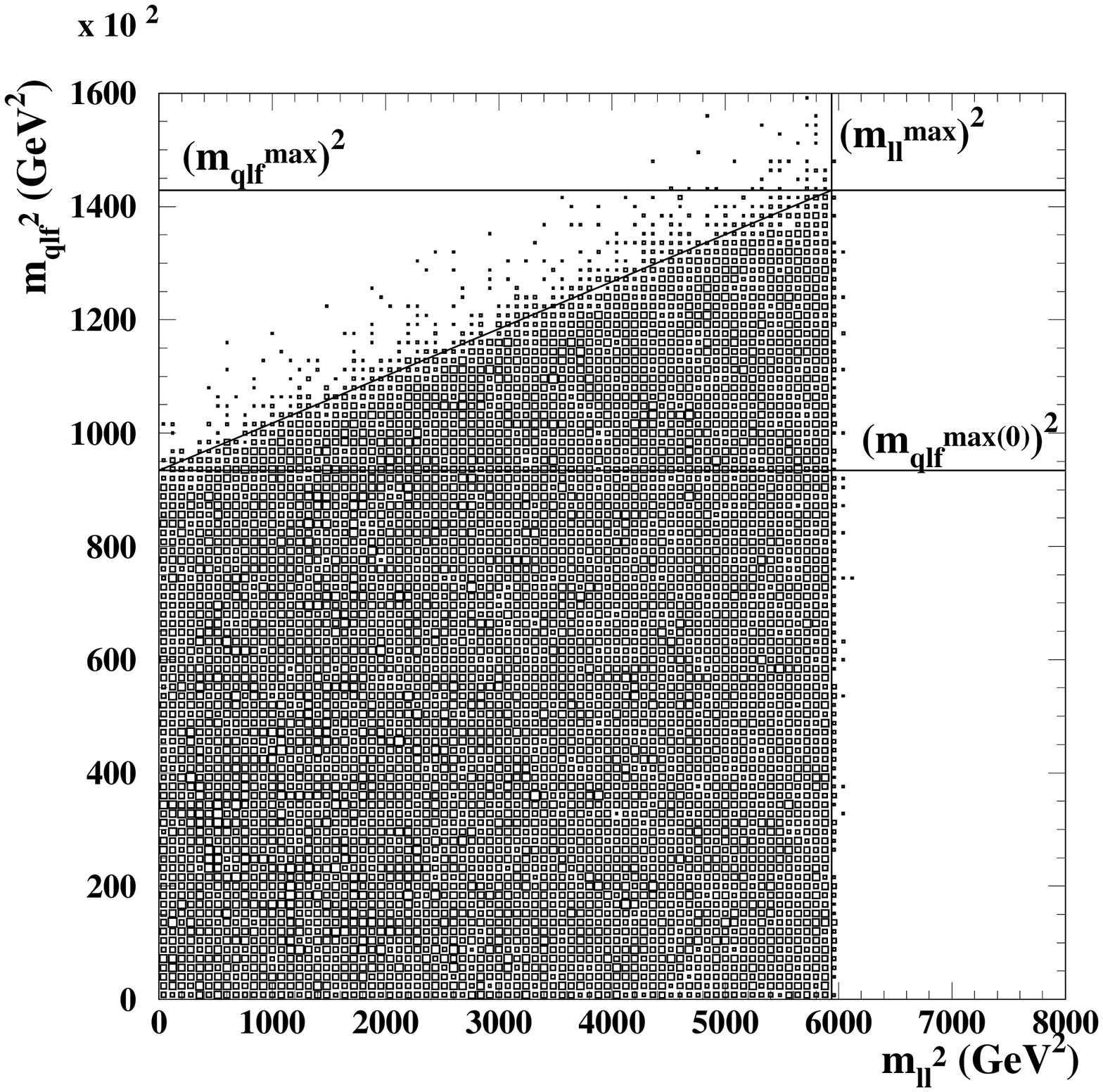,height=2.5in}
\epsfig{file=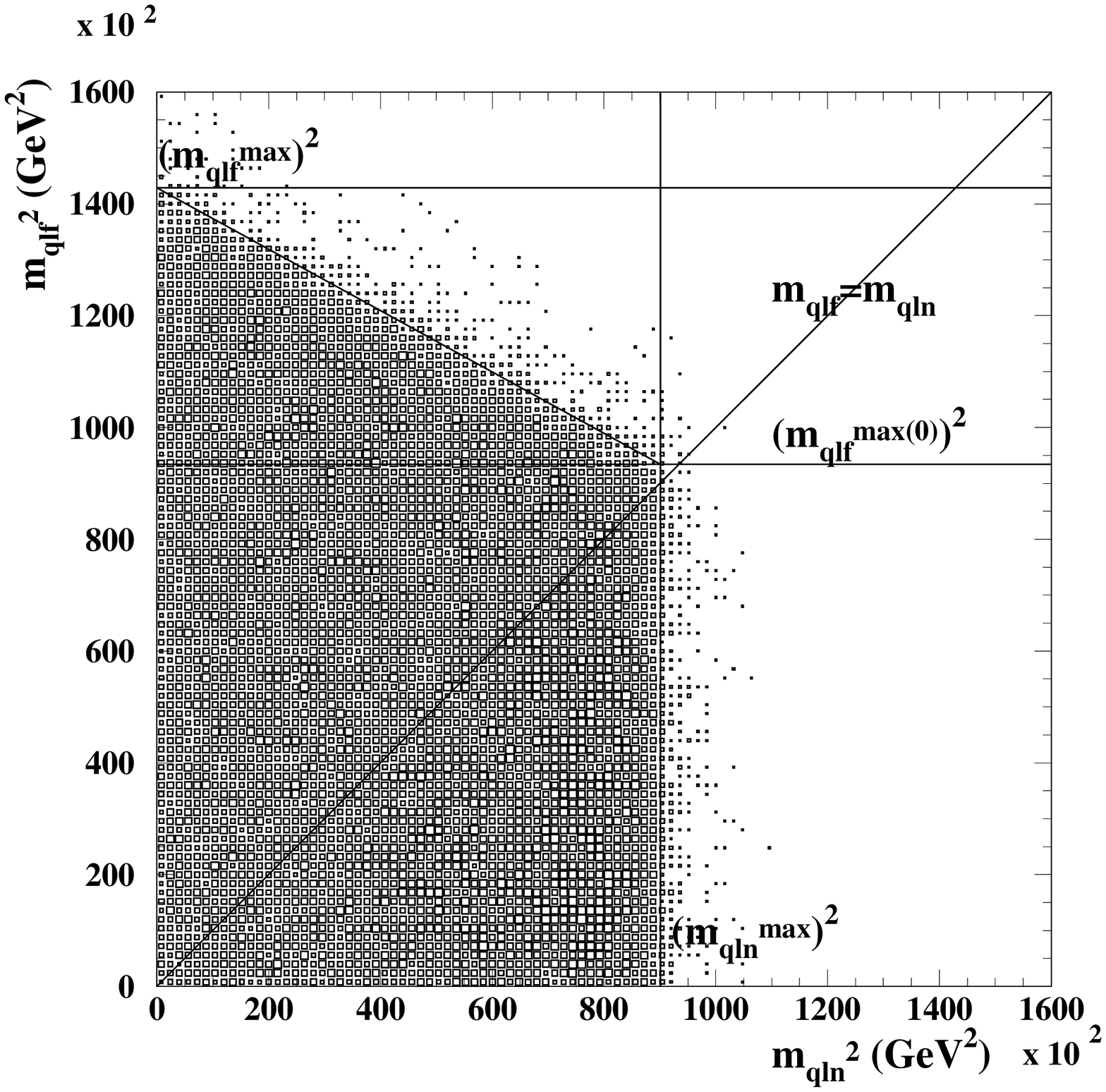,height=2.5in}
\caption{\label{fig2} Parton-level two-dimensional invariant
mass-squared distributions for decay chain Eqn.~(\ref{eqn2}) for the
SPS1a benchmark model. The top-left figure shows the distribution of
$\mqln^2$ as a function of $\mll^2$, the top-right figure the
distribution of $\mqlf^2$ as a function of $\mll^2$ and the bottom
figure the distribution of $\mqlf^2$ as a function of $\mqln^2$. See
text for explanation of bounds.} }
\FIGURE[ht]{ 
\epsfig{file=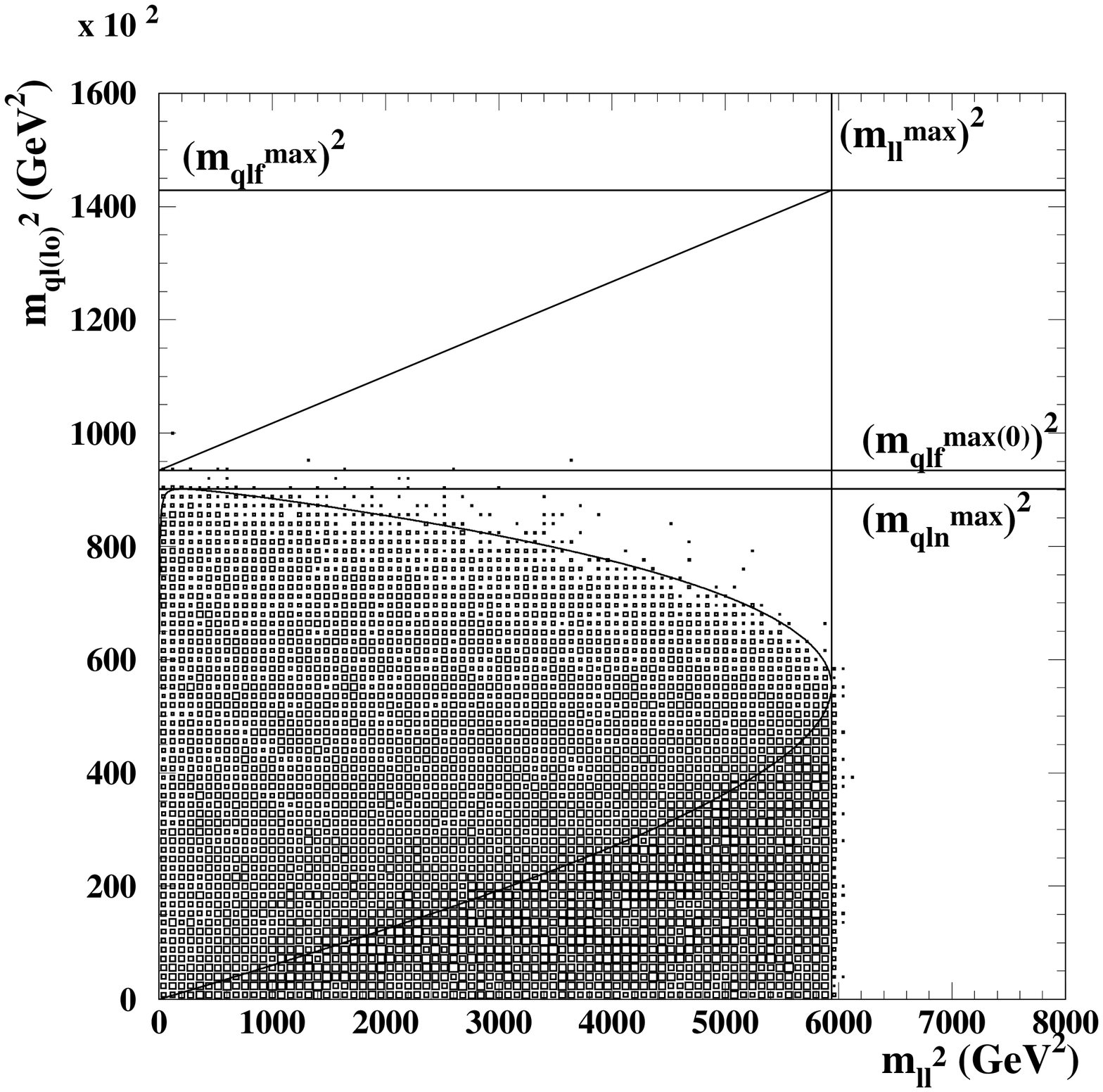,height=2.5in}
\epsfig{file=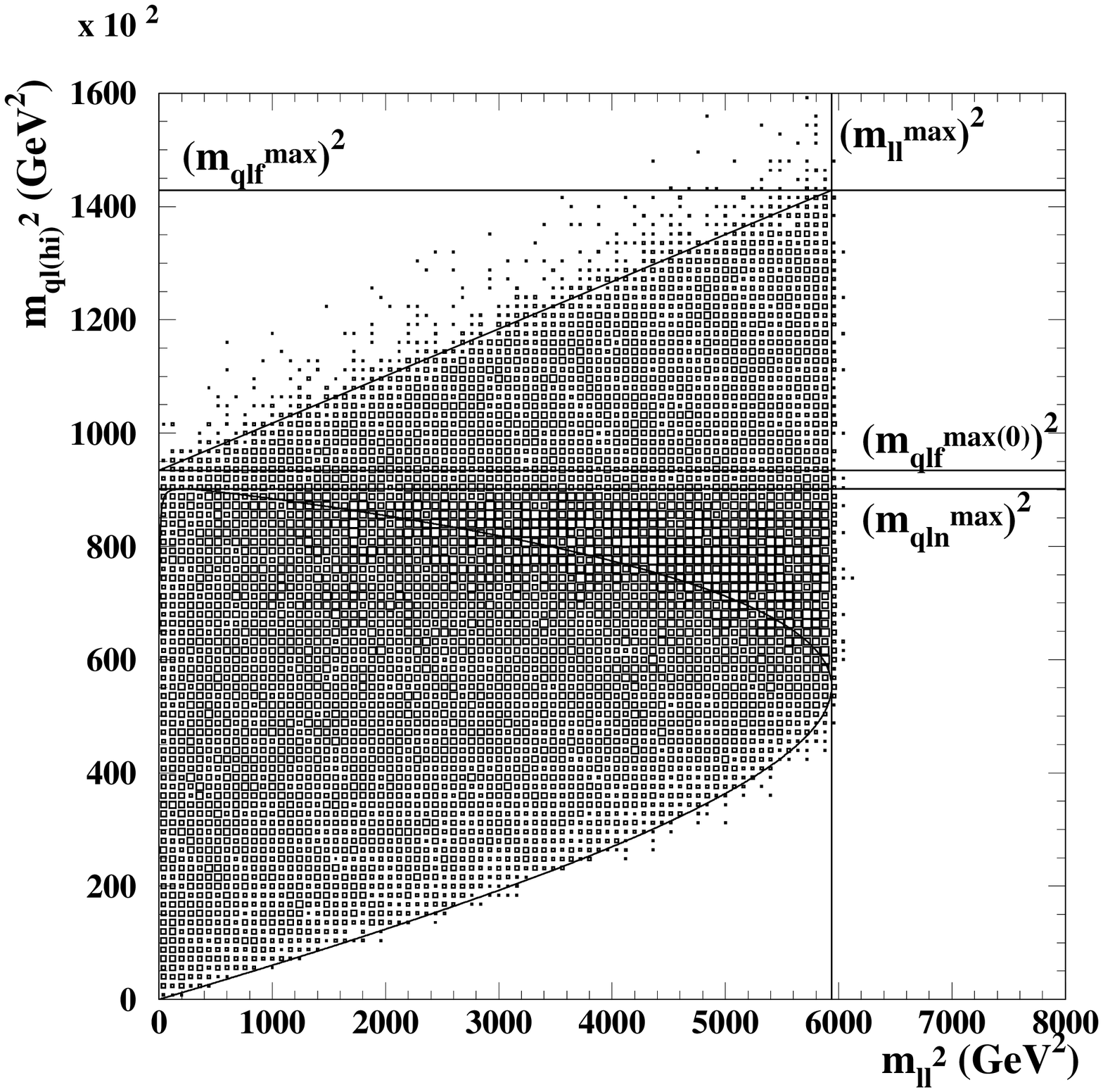,height=2.5in}
\epsfig{file=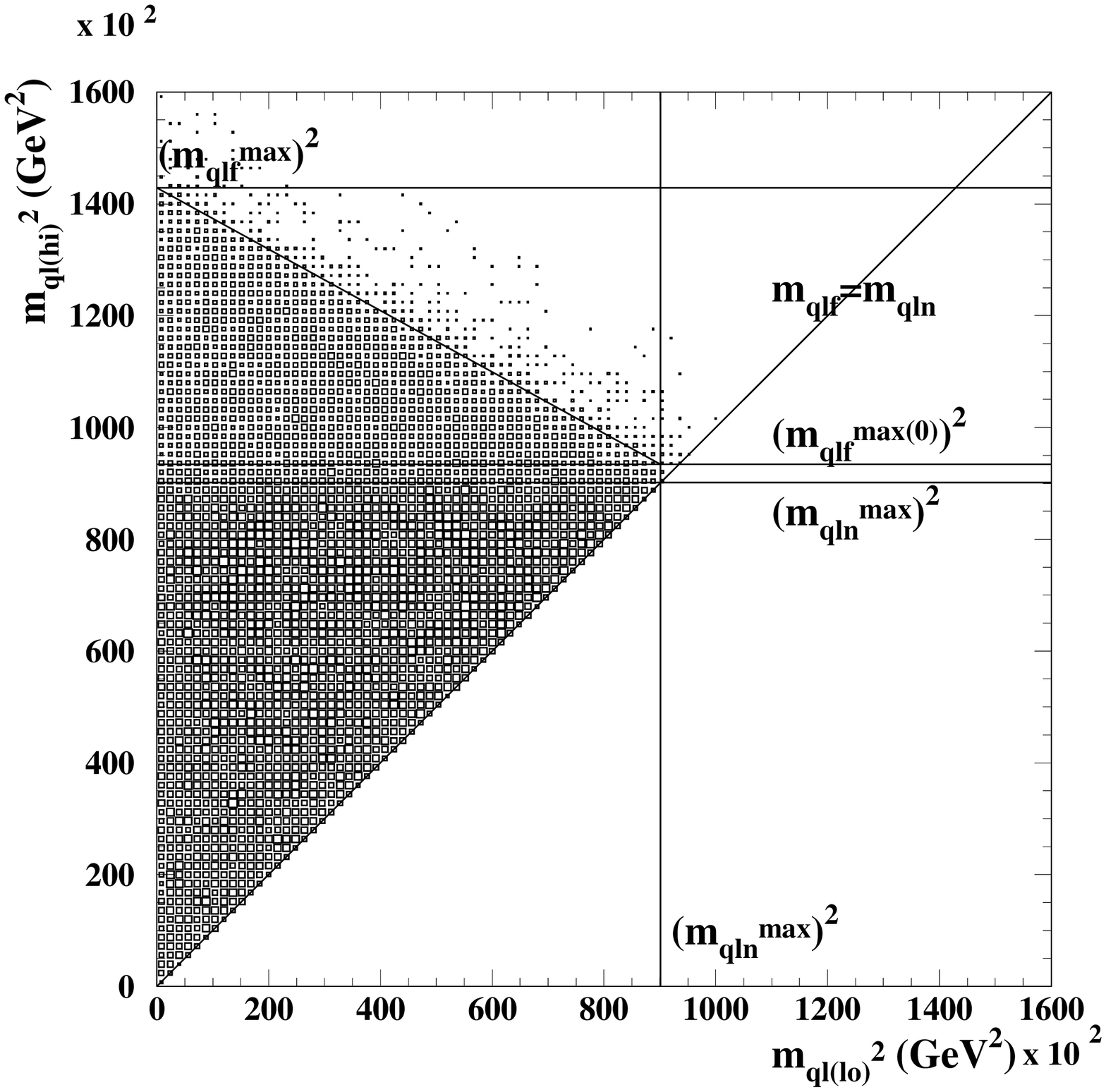,height=2.5in}
\caption{\label{fig3} Observable two-dimensional parton-level
invariant mass-squared distributions for decay chain Eqn.~(\ref{eqn2})
for the SPS1a benchmark model. The top-left figure shows the
distribution of $\mqllo^2$ as a function of $\mll^2$, the top-right
figure the distribution of $\mqlhi^2$ as a function of $\mll^2$ and
the bottom figure the distribution of $\mqlhi^2$ as a function of
$\mqllo^2$. See text for explanation of bounds.} }

In Figure~\ref{fig2} the bounds on $\mqlf^2$, $\mqln^2$ and $\mll^2$
obtained from the conventional `global end-point' analysis
together with $\big(\mqlfmaxmllzero\big)^2$ are illustrated by
vertical and horizontal lines (labelled). In the top-right and bottom
figures the upper bounds on $\mqlf^2$ lie respectively at:
\begin{equation}
\label{eqn29a}
\big(\mqlfmaxmll\big)^2 = \big(\mqlfmaxmllzero\big)^2 + \left[\big(\mqlfmax\big)^2-\big(\mqlfmaxmllzero\big)^2\right]\left(\frac{\mll}{\mllmax}\right)^2,
\end{equation}
and 
\begin{equation}
\label{eqn29b}
\big(\mqlfmaxmqln\big)^2 = \big(\mqlfmax\big)^2 - \left[\big(\mqlfmax\big)^2-\big(\mqlfmaxmllzero\big)^2\right]\left(\frac{\mqln}{\mqlnmax}\right)^2.
\end{equation}

Of course in practice we can not distinguish $\mqlf$ from $\mqln$ on
an event-by-event basis. For this reason we plot in Figure~\ref{fig3}
the equivalent distributions to Figure~\ref{fig2} for $\mqllo^2$ and
$\mqlhi^2$ rather than $\mqln^2$ and $\mqlf^2$. As expected the
distributions resemble combinations of the $\mqln^2$ and $\mqlf^2$
distributions, with an additional bound where
$\mqln=\mqlf=\mqleq$. This bound is trivial in
Figure~\ref{fig3}(bottom) while in the top-left and top-right figures
the equation of the bound can be obtained by using
Eqn.~(\ref{eqn26}) and Eqns.~(\ref{eqn14})--(\ref{eqn16}) and solving
the resulting quadratic equation for $\mqleq^2$ as a function of
$\mll^2$:
\begin{equation}
\label{eqn100}
\big(\mll^2+A^2\big) \mqleq^4 - 2\mll^2p_q\big(2p_{l_n}+A\big) \mqleq^2 + \mll^4 p_q^2 = 0,
\end{equation}
where
\begin{equation}
A \equiv rp_{l_f}'-p_{l_n}+\big(1-r^2\big)\frac{\mll^2}{4p_{l_n}}.
\end{equation}
The two roots of this equation give the upper and lower curved
bounds in these figures.

\clearpage

\subsection{Conditional end-point constraints with three-dimensional correlations}
\label{sec4.3}
A second class of constraints can be obtained from Eqn~(\ref{eqn26})
by measuring invariant mass correlations without integration over
additional mass distributions. In principle this could be used to measure
$\Sigma_{qll}$ directly. Here we provide one example of such a
technique in which we study the form of the one-dimensional lines
where $\Sigma_{qll}$ intersects the planes with respectively
$x$, $y$ and $z$ (or equivalently $\mqln$, $\mqlf$
and $\mll$) equal to zero. The resulting mass-squared distributions
are analogous to those shown in Figures~\ref{fig2} and~\ref{fig3} but
with the additional requirement that the third mass(-squared) value
not plotted in each figure is small.
\FIGURE[ht]{ \epsfig{file=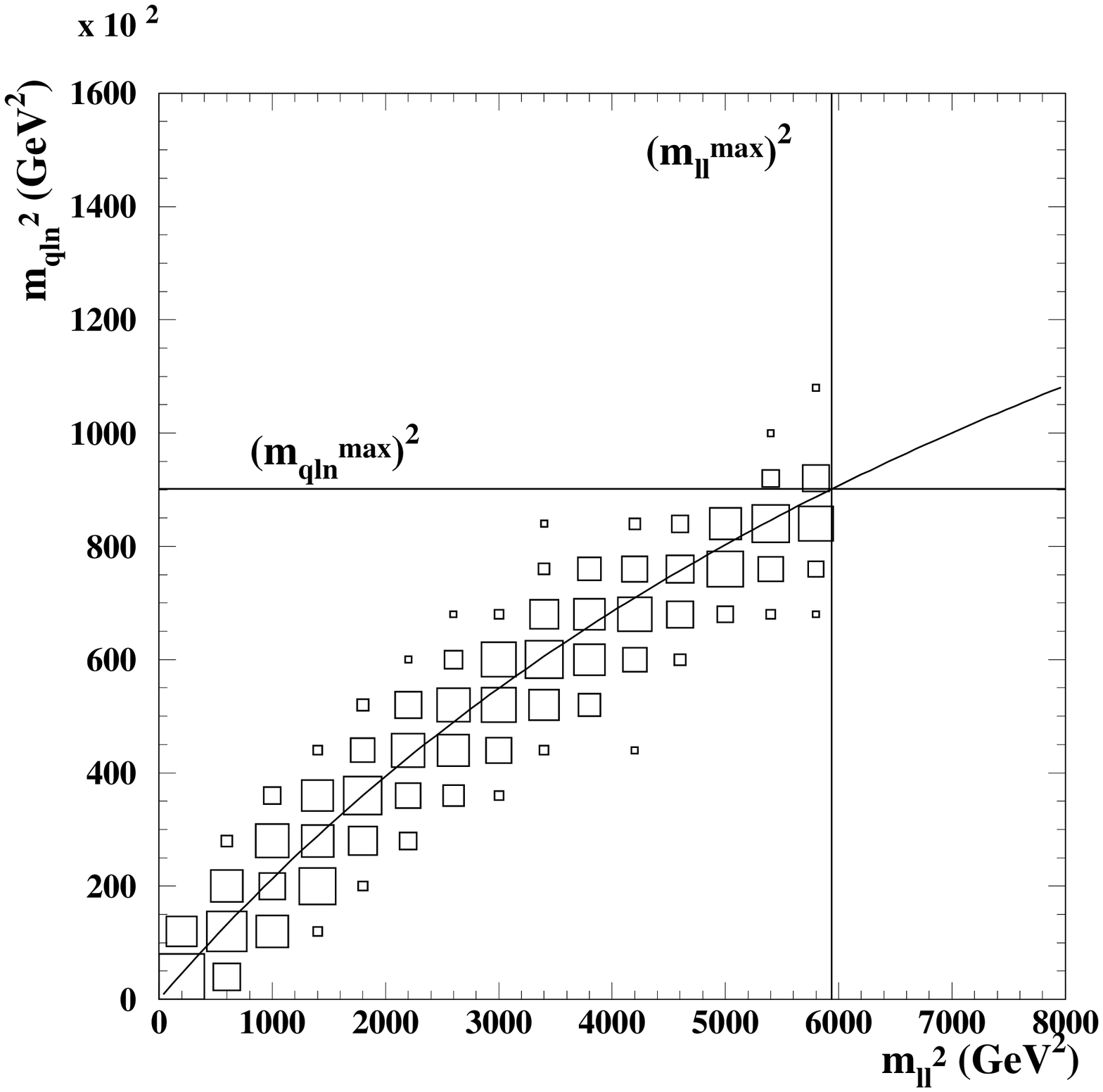,height=2.5in}
\epsfig{file=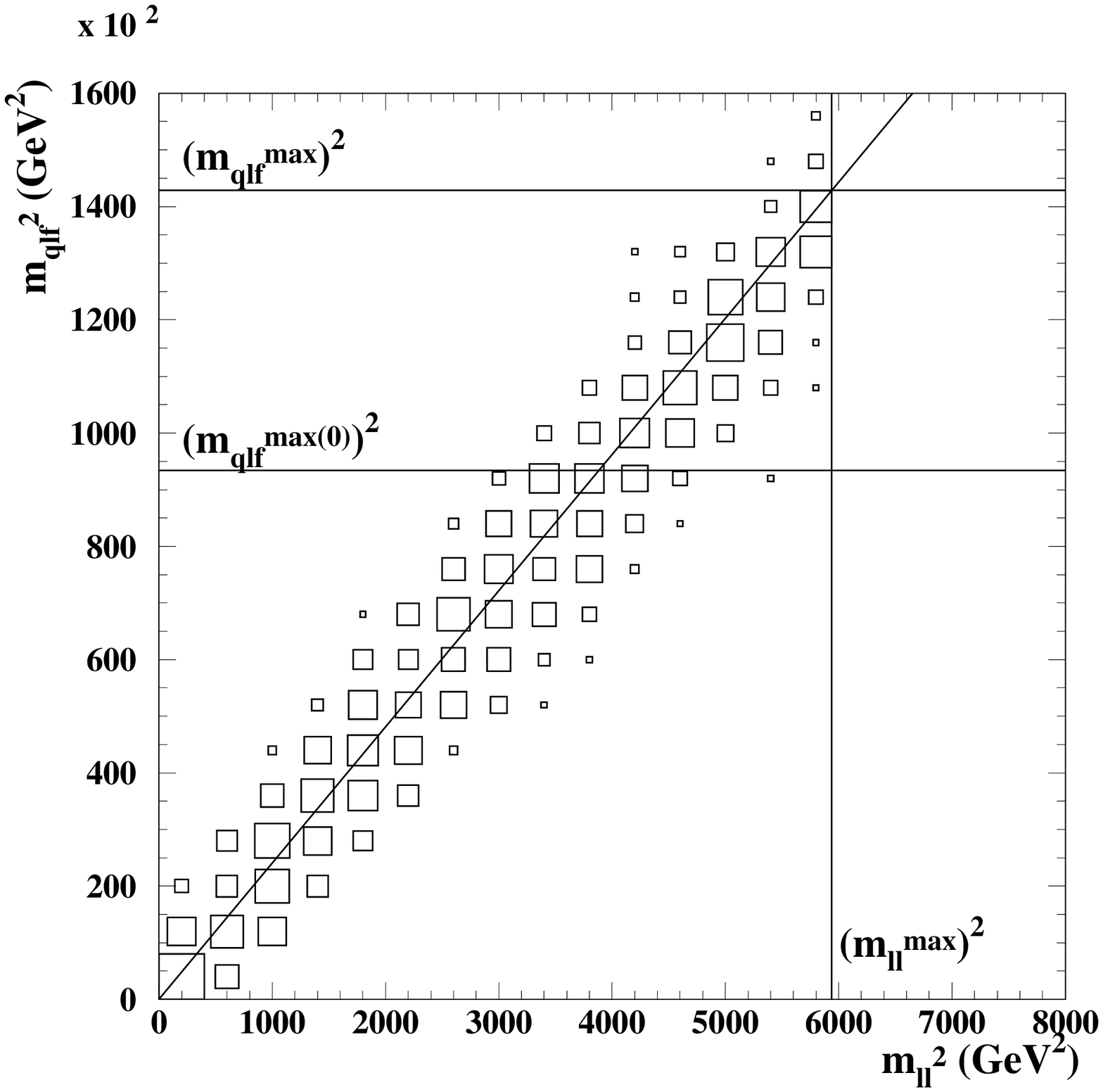,height=2.5in}
\epsfig{file=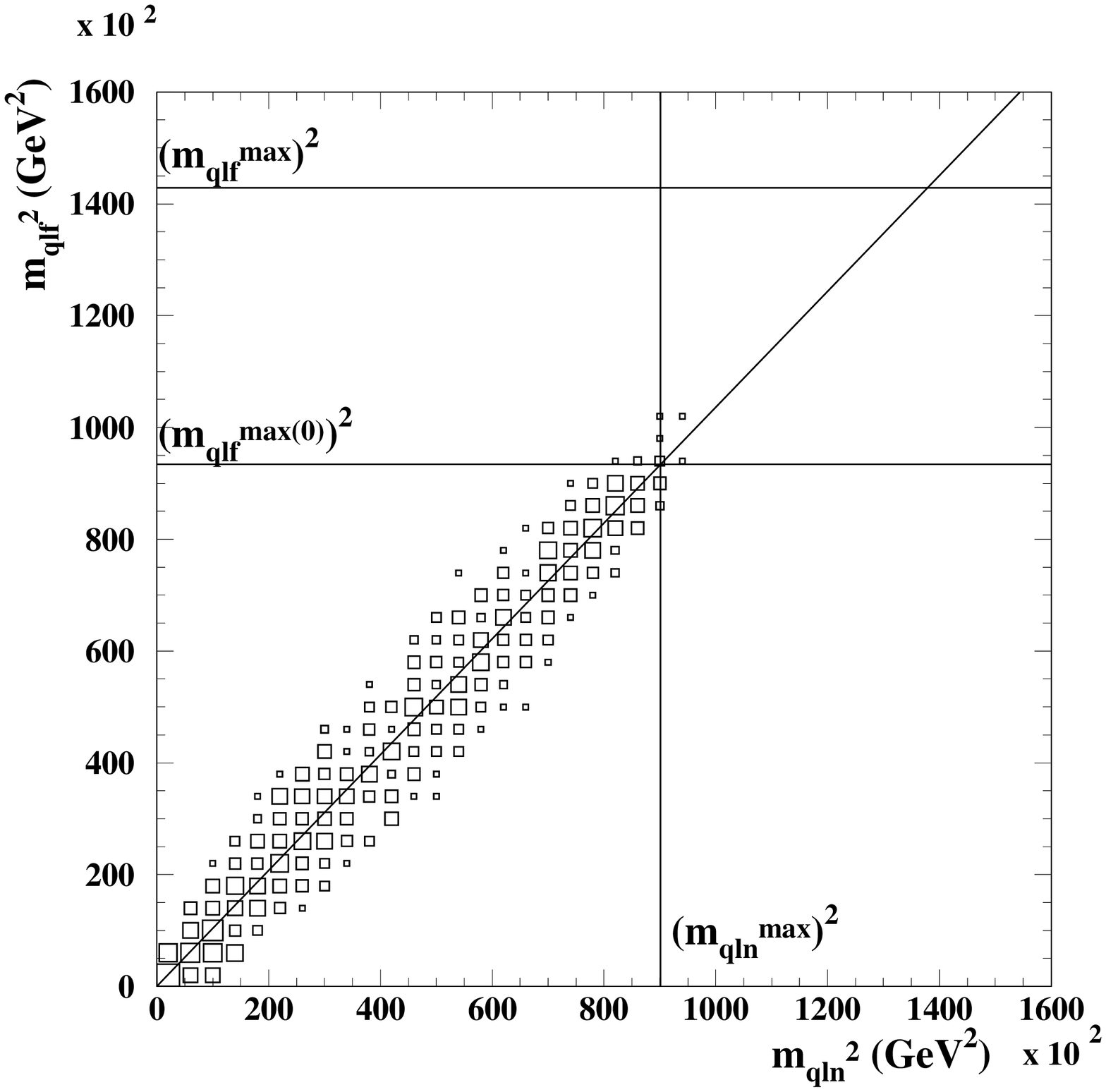,height=2.5in}
\caption{\label{fig4} Two-dimensional distributions of two-particle
invariant masses for the SPS1a model, with the third invariant mass
required to be less than 50 GeV (top figures) or less
than 10 GeV (bottom figure). The top-left figure shows $\mqln^2$
versus $\mll^2$, the top-right figure $\mqlf^2$ versus $\mll^2$, and
the bottom figure $\mqln^2$ versus $\mqlf^2$. See text for explanation
of bounds and loci.} }
\FIGURE[ht]{
\epsfig{file=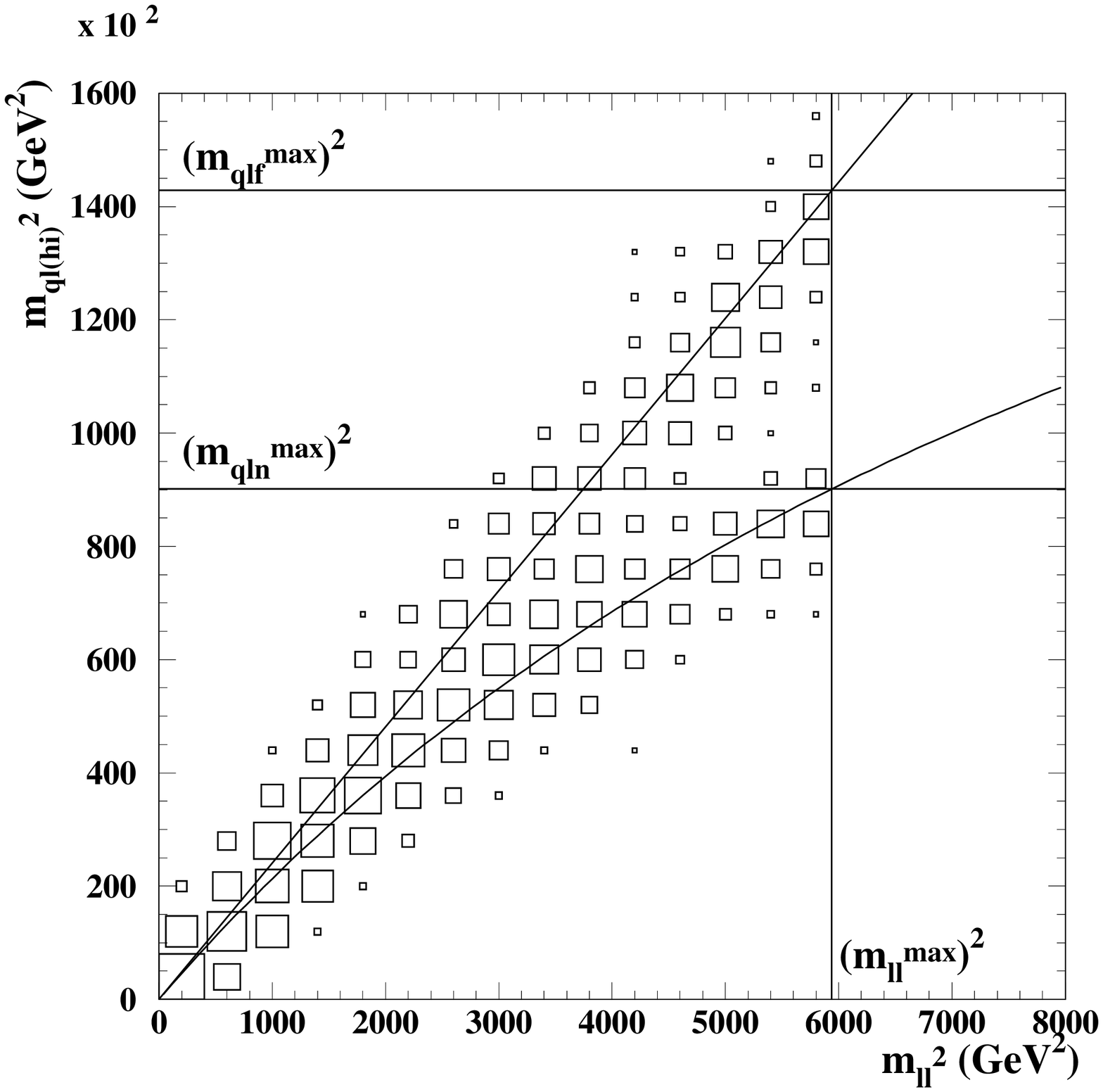,height=2.5in}
\epsfig{file=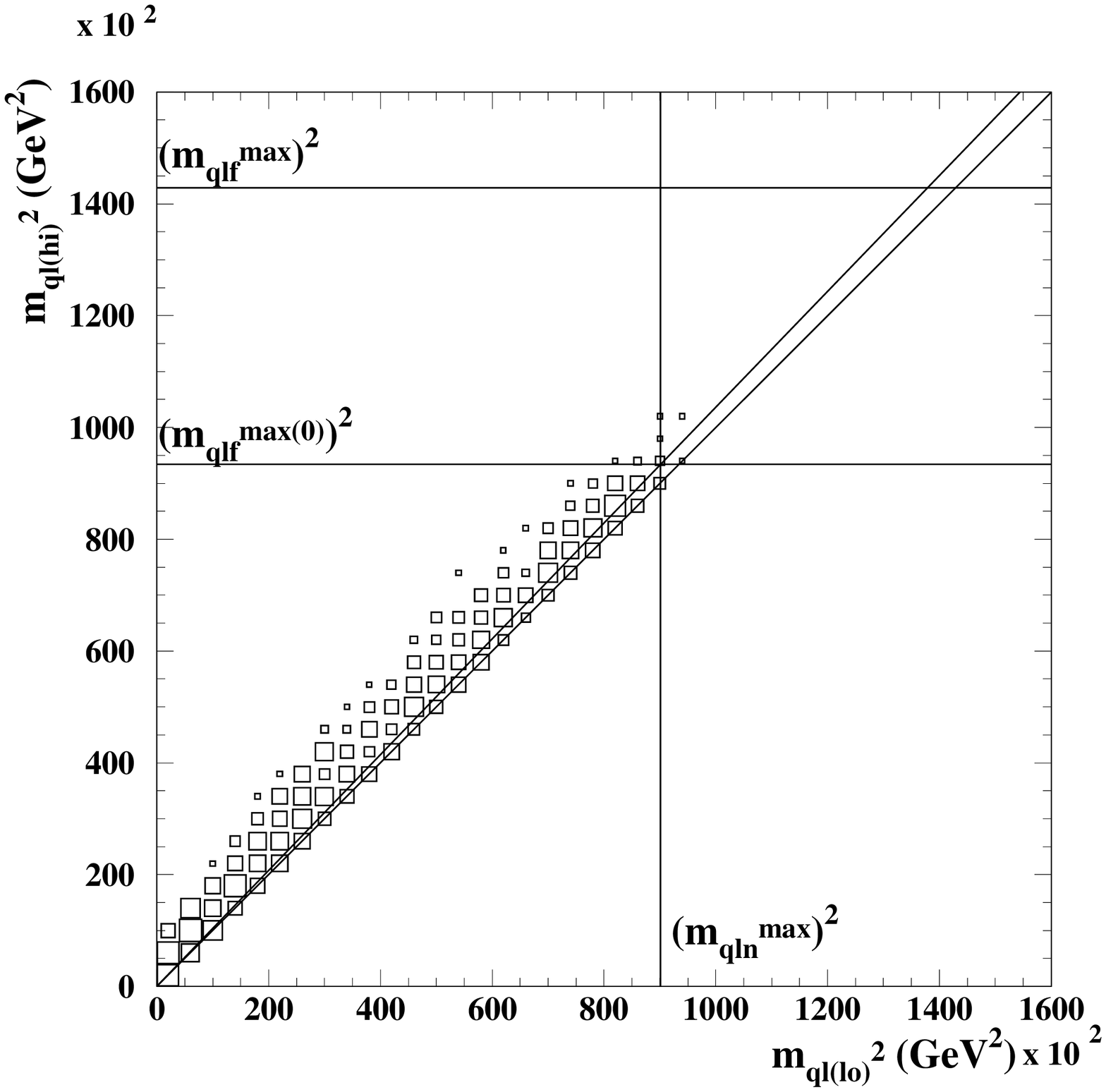,height=2.5in}
\caption{\label{fig5} Observable distributions constructed from the
quantities plotted in Figure~\ref{fig4} for the SPS1a benchmark
model. The lower of the two diagonal lines in the right-hand figure
represents the trivial bound where $\mqlhi=\mqllo$. The remaining
bounds and loci are as for Figure~\ref{fig4} (see text).} }

We proceed by requiring $\mqln$, $\mqlf$ or $\mll$ to be less than
some value (here set to 50 GeV for $\mql$ quantities and 10 GeV for
$\mll$)\footnote{With these selection requirements the mean
quark-lepton(lepton-lepton) separation is $\Delta R=0.6(0.2)$. If
these object separations were too small in practice to provide useful
lepton reconstruction efficiency they could be increased at the
expense of mass resolution (see text).}  sufficiently large to provide
useful statistics but sufficiently small to approximately maintain the
one-dimensional form of the intersection. We then plot the
two-dimensional distribution of the two unconstrained squared masses
in each case. These distributions can be seen in Figure~\ref{fig4}. In
the top-left and top-right figures the vertical lines represent the
bound from $\big(\mllmax\big)^2$, while the horizontal lines represent
the bounds from $\big(\mqlnmax\big)^2$, $\big(\mqlfmax\big)^2$ and
$\big(\mqlfmaxmllzero\big)^2$. In the bottom figure the vertical and
horizontal lines represent the bounds from $\big(\mqlnmax\big)^2$ and
$\big(\mqlfmaxmllzero\big)^2$ respectively. The loci of the
distributions in the figures are given in terms of dimensionless mass
coordinates by:
\begin{eqnarray}
x^2(y=0) & = & \frac{z^2}{r+\frac{1}{4}\left(1-r^2\right)z^2},\label{eqn34}\\
y^2(x=0) & = & z^2, \label{eqn35}\\
y^2(z=0) & = & x^2r, \label{eqn36}
\end{eqnarray}
which gives the following relations:
\begin{eqnarray}
\mqln^2 & = & \mll^2 \left(\frac{\msql^2-\mchioii^2}{\mslr^2-\mchioi^2+\mll^2}\right),  \label{eqn37} \\
\mqlf^2 & = & \mll^2 \left(\frac{\msql^2-\mchioii^2}{\mchioii^2-\mslr^2} \label{eqn38}\right), \\
\mqlf^2 & = & \mqln^2 \left(\frac{\mslr^2-\mchioi^2}{\mchioii^2-\mslr^2}\right). \label{eqn39}
\end{eqnarray}
\FIGURE[ht]{ \epsfig{file=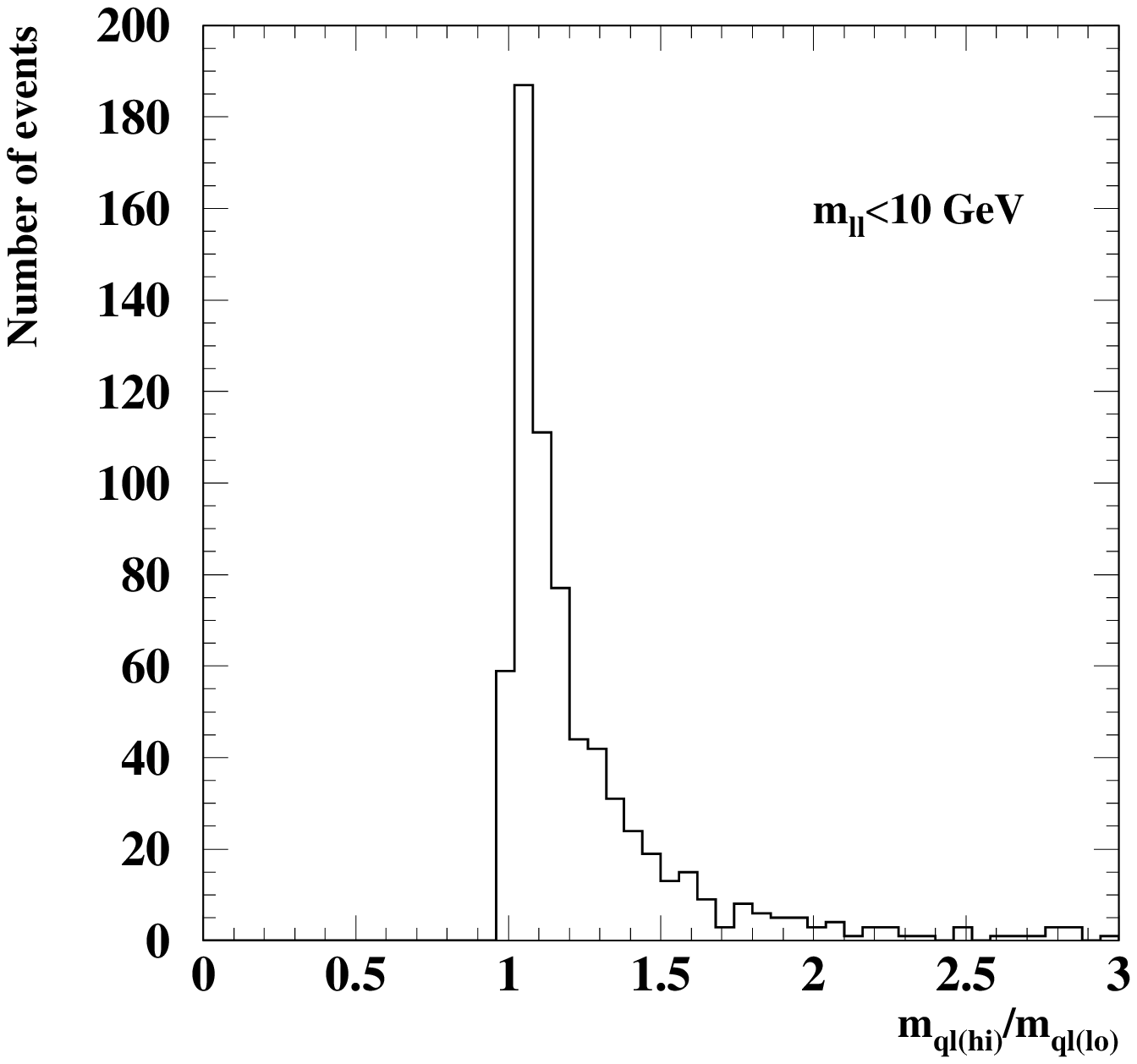,height=3.5in}
\caption{\label{fig6} Distribution of the ratio $\mqlhi/\mqllo$ for
$\mll$ $<$ 10 GeV for the SPS1a benchmark SUSY model. The distribution
possesses a trivial bound at $\mqlhi/\mqllo=1$ while the prominent
peak is indicative of sequential two-body lepton producing decays. } }

The distributions in Figure~\ref{fig4} are not observable directly due
to the $l_n/l_f$ ambiguity. Instead the three distributions of
Figure~\ref{fig4} are replaced by two distributions (with $\mqllo$
required to be small), plotted in Figure~\ref{fig5}. If the loci in
this figure can be identified in the presence of realistic
detector-level smearing then they could be used to measure sparticle
masses via Eqns.~(\ref{eqn37})--(\ref{eqn39}). In addition however
Figure~\ref{fig5}(right) can be used to determine whether the decay
chain in selected events consists of two sequential two-body
lepton-producing decays (as in Eqn.~(\ref{eqn2})) or a single three-
(or more) body decay (e.g. $\chioii \rightarrow l^+l^-\chioi$). In the
latter case $\mll$ can acquire a small value through one of the two
leptons acquiring a very small momentum in the $\chioii$ rest frame,
while in the former case this can only occur due to event topology,
i.e. due to the two leptons being emitted co-linearly in this
frame. This conclusion may also be reached by observing that
$\Sigma_{qll}$ defined in Eqn.~(\ref{eqn26}) does not intersect any of
the $x$, $y$ or $z$ coordinate axes and instead intersects the $x=0$,
$y=0$ and $z=0$ planes in one-dimensional lines generating the loci
shown in Figure~\ref{fig4}. The consequence of this observation is
that in the three-body case $\mqlhi$ and $\mqllo$ should be rather
less correlated than in the sequential two-body case, where $\mqlhi$
is linearly related to $\mqllo$ as shown in
Figure~\ref{fig5}(right). A convenient means of observing this
correlation is to plot the one dimensional distribution of the ratio
$\mqlhi/\mqllo$ shown in Figure~\ref{fig6}, where the presence of the
prominent peak provides evidence for the presence of sequential
two-body lepton producing decays. This provides a useful alternative
to conventional techniques such as studying the shape of the $\mll$
distribution or measuring the ratio $\mqlhimax/\mqllomax$ (which
equals $\sqrt{2}$ for three-body lepton-producing decays
\cite{Lester:2006cf}). The kinematic bounds for three-body
lepton-producing decay chains incorporating single two-body
quark-producing decays are listed for completeness in Appendix A.

\section{Mass reconstruction}
\label{sec5}
Having identified some correlations between invariant mass
observables we shall now discuss how they might be used in principle
to constrain individual sparticle masses.

We start by observing that measurements of the global end-points
together with measurements of the straight-line upper bounds in the
two-dimensional distributions shown in Figure~\ref{fig3} can in
principle give direct access to the four `core' two-body observables
$\mllmax$, $\mqlnmax$, $\mqlfmax$ and $\mqlfmaxmllzero$. In order to
measure these quantities however we must first resolve or evade the
ambiguities in the interpretation of the $\mql$ and $\mqll$ end-points
listed in Eqns.~(\ref{eqn6}) and~(\ref{eqn8}). This can be
accomplished by examining the form of the kinematic bounds in
Figures~\ref{fig2} and~\ref{fig3}. The upper $\mqln^2$ bound (top-left
panel of Figure~\ref{fig2}) is constant as a function of $\mll^2$,
while the equivalent $\mqlf^2$ bound (top-right panel of
Figure~\ref{fig2}) is a linear function of $\mll^2$ with non-zero
gradient (in the absence of sparticle mass degeneracies). Consequently
if the conditional upper bound on $\mqlhi$ changes with $\mll$ then
$\mqlhimax=\mqlfmax$, corresponding to cases A1 and A2 in
Eqn.~(\ref{eqn6}). This is the situation for SPS1a. Conversely if the
upper bound on $\mqlhi$ remains constant with $\mll$ then
$\mqlhimax=\mqlnmax$, which corresponds to case A3 in
Eqn.~(\ref{eqn6}), and hence also $\mqllomax=\mqleqmax$.

In order to discriminate between cases A1 and A2 we could in principle
examine the behaviour of the conditional upper bound on $\mqllo$ as a
function of $\mll$. If the bound is constant in the vicinity of
$\mqllomax$ then $\mqllomax=\mqlnmax$ (case A1, realised at SPS1a),
while if it changes as a function of $\mll$ then $\mqllomax=\mqleqmax$
(case A2). Equivalently, if the distribution of $\mqlhi$ values as a
function of $\mqllo$ (shown for squared masses in
Figure~\ref{fig3}(bottom)) is bounded in $\mqllo$ by a vertical line
then $\mqllomax=\mqlnmax$ (case A1) but if it is bounded by a point
then $\mqllomax=\mqleqmax$ (case A2). Nevertheless the (squared mass)
distributions shown in Figure~\ref{fig3} for SPS1a, for which
$\mqlnmax$ is very close to $\mqleqmax$, illustrate the difficulty of
this strategy for resolving the $\mqllomax$ ambiguity. It is far from
clear in these parton-level distributions whether these conditions are
satisfied and at detector-level the difficulties will be still
greater. In this case however we can potentially also make use of the
fact that the upper bound on $\mqln^2$ shown in
Figure~\ref{fig2}(top-left) is also visible in the top-right and
bottom panels of Figure~\ref{fig3} as a discontinuity in the
$\mqlhi^2$ distribution below its upper bound. Consequently
observation of a `hidden' bound in the $\mqlhi$ distribution, with a
position independent of any conditions imposed upon $\mll$ would
enable $\mqlnmax$ to be measured directly, evading the A1/A2
ambiguity.

In cases A1 (e.g. SPS1a) or A2 the conditional upper bound on $\mqlhi$
as a function of $\mll$ allows $\mqlfmax$ and $\mqlfmaxmllzero$ to be
determined with Eqns.~(\ref{eqn29a}). Together with the measurement of
$\mqlnmax$ discussed above and the measurement of the $\mllmax$
end-point all four core two-body observables can hence be accessed. In
case A3 it is the upper bound on $\mqlf$ which is hidden but
nevertheless it may still be observable below the upper bound on
$\mqlhi$ provided here by the upper bound on $\mqln$. If this is the
case then all four core two-body observables can again be
determined. There are certainly special cases where the sparticle mass
values conspire to prevent ambiguity resolution on the basis of these
arguments, however in those cases the use of multiple redundant
measurements, for instance bounds on $\mqll$, should help to resolve
these ambiguities (see e.g. Ref.~\cite{Miller:2005zp}). The $\mll$
dependence of the conditional upper or lower bounds on $\mqlhi$ and
$\mqllo$ (shown for squared masses in Figure~\ref{fig3}) could also be
exploited.

An additional input to the mass reconstruction can be provided by the
global upper bound on $\mqll$ provided by $\mqllmax$. This bound also
suffers from ambiguities, as listed in Eqn.~(\ref{eqn8}). Having
resolved the ambiguities in $\mql$ however these ambiguities are
considerably easier to address. The three cases corresponding to
co-linear kinematic configurations (B1--B3) possess
$\big(\mqllmax\big)^2$ values of respectively
$\big(\mqlfmaxmllzero\big)^2+\big(\mqlnmax\big)^2$,
$\big(\mllmax\big)^2+\big(\mqlnmax\big)^2$, and
$\big(\mqlfmax\big)^2+\big(\mllmax\big)^2$. Consequently the position
of the value of $\mqllmax$ is already uniquely specified by
independent measurements of the four core two-body observables. If
$\big(\mqllmax\big)^2$ does not equal any of these three values within
errors then case B4 is likely to provide the bound and hence an
additional kinematic constraint. It should be noted that further mass
constraints can in principle be obtained using Eqn.~(\ref{eqn8}) by
measuring the dependence of the conditional upper bound on $\mqll$ as
a function of $\mll$ together with the similar dependence of the
conditional lower bound on $\mqll$, from which the global $\mqllmin$
end-point derives.

If we can unambiguously measure at least the four core two-body
observables then we can obtain the individual sparticle masses
analytically from:
\begin{eqnarray}
\mchioi &=& \frac{\mllmax\mqlfmaxmllzero}{\big(\mqlfmax\big)^2-\big(\mqlfmaxmllzero\big)^2}\sqrt{\big(\mqlnmax\big)^2-\big(\mqlfmax\big)^2+\big(\mqlfmaxmllzero\big)^2}, \label{eqn101} \\
\mslr &=& \frac{\mllmax\mqlnmax\mqlfmaxmllzero}{\big(\mqlfmax\big)^2-\big(\mqlfmaxmllzero\big)^2}, \label{eqn102} \\
\mchioii &=& \frac{\mllmax\mqlnmax\mqlfmax}{\big(\mqlfmax\big)^2-\big(\mqlfmaxmllzero\big)^2}, \label{eqn103} \\
\msql &=& \frac{\mqlnmax\mqlfmax}{\big(\mqlfmax\big)^2-\big(\mqlfmaxmllzero\big)^2}\sqrt{\big(\mllmax\big)^2+\big(\mqlfmax\big)^2-\big(\mqlfmaxmllzero\big)^2}. \label{eqn104}
\end{eqnarray}
Additional constraints, for instance provided by $\mqllmax$,
$\mqllmin$ or measurements of the conditional upper or lower bounds on
$\mqlhi$ and $\mqllo$ can be included in several equivalent ways. One
approach involves replacing the analytical formulae of
Eqns.~(\ref{eqn101})--(\ref{eqn104}) with a numerical fit to the
sparticle masses incorporating all constraints. This approach is
discussed further in Section~\ref{sec6}. An alternative procedure
involves re-casting the constraint equations in terms of the four core
two-body observables and using a numerical fit to determine these
quantities, retaining the analytical formulae for determining the
sparticle masses in a second step. In this approach
Eqns.~(\ref{eqn32}) and~(\ref{eqn100}) can be re-cast in terms of
$\mqlnmax$, $\mqlfmax$, $\mqlfmaxmllzero$ and $\mllmax$ with
Eqns.~(\ref{eqn101})--(\ref{eqn104}) together with the equations
relating these quantities to the momenta of the jets and leptons
considered in Section~\ref{sec3}:
\begin{eqnarray}
p_{l_n} &=& \frac{\mllmax\mqlnmax}{2\mqlfmax},\label{eqn105} \\
p_q &=& \frac{\mqlnmax\mqlfmax}{2\mllmax}, \label{eqn106} \\
p_{l_f}' &=& \frac{\mqlfmaxmllzero\mllmax}{2\mqlnmax}. \label{eqn107}
\end{eqnarray}

\section{Detector-level study}
\label{sec6}

\subsection{Introduction}
\label{sec6.1}
Having outlined some possible sparticle mass measurement strategies
making use of the additional information provided by invariant mass
correlations we now illustrate how such strategies may be used in
practice with detector-level events. Although a very wide range of new
variables are potentially now available we shall concentrate on a
small subset representing a limited evolution of the conventional
global end-point technique. This will enable us to highlight the
main benefit of the new technique, namely the amiguity resolution
discussed in Section~\ref{sec5}, without compromising event statistics
near the end-points.

We shall focus purely on $\mqll$ and $\mql$ end-points as functions of
$\mll$ in order to minimise the possibility of under-estimating
statistical uncertainties through the presence of correlations between
end-point observables. Such correlations could arise if the same
events appear at the end-points of two or more invariant mass
distributions. Estimation of these correlations is beyond the scope of
the current paper and so we choose instead to minimise their effects
by judicious selection of minimally correlated observables. With more
work to understand such correlations further improvements in mass
measurement precision could potentially be obtained.

It should be noted that throughout this section we consider end-points
in distributions of invariant masses rather than squared invariant
masses. The use of the former here enables comparison of results with
those of previous studies, and is consistent with the discussion of
Section~\ref{sec5}.

\subsection{Event simulation, selection and reconstruction}
\label{sec6.2}
The 100 fb$^{-1}$ equivalent sample of SPS1a events described in
Section~\ref{sec2.1} was passed through a fast simulation of a generic
LHC detector \cite{RichterWas:2002ch}. In addition a 100 fb$^{-1}$
equivalent sample of $t\bar{t}$ background events was generated with
{\tt HERWIG 6.4} \cite{Corcella:2000bw}, passed through the same fast
detector simulation, filtered to require at least two leptons ($e$ or
$\mu$) and added to the signal sample. $t\bar{t}$ events are expected
to form the dominant SM background to the analysis described here,
however the event selection and background subtraction described below
are found to reduce this contribution to negligible levels (typically
$\lesssim$ 10 events for 100 fb$^{-1}$ of data).  

Events were selected with the same requirements as were used in
Ref.~\cite{Weiglein:2004hn} to aid comparison:
\begin{itemize}
\item At least four jets with the hardest three satisfying: $p_T(j_1)$
$>$ 150 GeV, $p_T(j_2)$ $>$ 100 GeV, $p_T(j_3)$ $>$ 50 GeV
\item $M_{\rm eff} = \ETM + p_T(j_1) + p_T(j_2) + p_T(j_3) +
p_T(j_4)$ $>$ 600 GeV
\item $\ETM$ $>$ $\max(100 {\rm GeV}, 0.2M_{\rm eff})$
\item Exactly two isolated opposite-sign same-flavour (OSSF) leptons
($e/\mu$) satisfying: $p_T(l_1)$ $>$ 20 GeV, $p_T(l_2)$ $>$ 10 GeV.
\end{itemize}
Following event selection the invariant mass of the OSSF lepton pair
($\mll$) was calculated for each event. The $\mll$ distribution of
events containing opposite-sign opposite-flavour (OSOF) lepton pairs
was subtracted from that of OSSF events in order to remove SM and SUSY
background events containing uncorrelated leptons. The resulting
$\mll$ distribution is shown in Figure~\ref{fig7}(top). This OSOF
subtraction procedure was also applied when generating all of the
invariant mass distributions shown below. These distributions also all
require that $\mll$ $<$ 80 GeV (3 GeV above the observed value of
$\mllmax$).
\FIGURE[ht]{
\epsfig{file=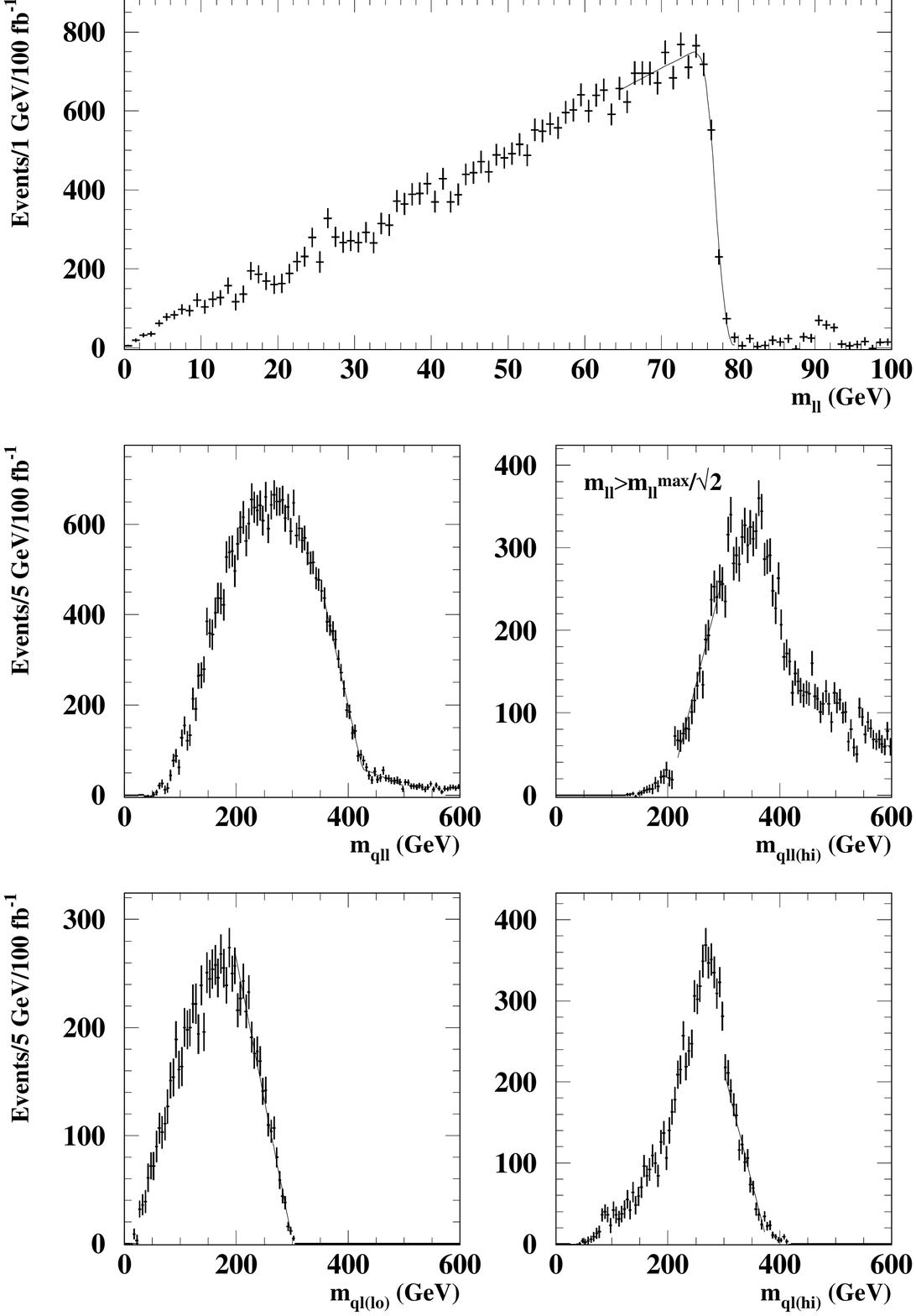,height=7.5in}
\caption{\label{fig7} Detector-level invariant mass distributions for
the SPS1a benchmark SUSY model used in the conventional integrated
end-point analysis.  } }

The next step in the analysis involves indentifying the hard jet
generated by the assumed decay chain Eqn.~(\ref{eqn2}). Throughout
this analysis this jet was assumed to be one of the two leading jets
in the event. When calculating the maxima of invariant mass
distributions this jet was further assumed conservatively to be the
one which minimises $\mqll$, while when calculating thresholds and
invariant mass ratios the jet which maximises $\mqll$ was used. This
latter $\mqll$ value is referred to as $\mqllhi$ below. When
calculating $\mql$ end-points only those events in which the two
possible $\mqll$ values lie either side of the observed $\mqllmax$
end-point were used. When calculating distributions of ratios of
invariant masses this requirement was strengthened to require that
both $\mqll$ values lie below the observed $\mqllmax$ end-point.

\subsection{Global end-point analysis}
\label{sec6.3}

The integrated $\mqll$, $\mqllhi$, $\mqllo$ and $\mqlhi$
distributions obtained from this analysis are shown in the bottom four
panels of Figure~\ref{fig7} and agree well with the results of earlier
work \cite{Weiglein:2004hn}. These distributions were fitted with
linear functions to determine the positions of the global end-points
while the $\mll$ distribution (Figure~\ref{fig7}(top)) was fitted with
a `triangular' distribution smeared with a gaussian resolution
function of variable width, as described in
e.g. Ref.~\cite{Allanach:2000kt}.  A more detailed analysis might
involve fitting with the analytical functions described in
Ref.~\cite{Miller:2005zp} or simulated detector-level template
distributions, however such developments are beyond the scope of this
paper.

The mean end-point positions and their associated statistical
uncertainties obtained from the fits are shown in
Table~\ref{tab1}. The agreement between the nominal end-point positions
expected from the masses used in the {\tt HERWIG} generator and the
fitted means is reasonable in all cases. Also shown in
Table~\ref{tab1} are the jet/lepton energy scale (JES and LES)
systematic uncertainties, assumed to be 1\% and 0.1\%
respectively. The scale uncertainties in the $\mql$ and $\mqll$
end-points are dominated by the JES uncertainty leading to a 100\%
correlation between the systematic uncertainties in these
observables. As discussed in Ref.~\cite{Weiglein:2004hn} the
systematic uncertainties arising from variation of input parameters in
these naive fits are potentially significant ($\lesssim$10 GeV)
however for the purposes of this analysis we shall henceforth neglect
these systematics.
\TABLE[ht]{\small%
\begin{tabular}{|c|c|c|c|c|}
\hline
Observable &Input &Mean &Statistical error &Energy scale error\\
\hline
$\mllmax$  &77.053  &77.006  &0.057  &0.077 \\ 
$\mqllmax$ &428.4   &425.3   &1.8    &4.3 \\
$\mqllmin$ &201.8   &200.8   &3.1    &2.0 \\
$\mqlhimax$&377.9   &378.0   &1.6    &3.8 \\
$\mqllomax$&300.2   &301.3   &1.0    &3.0 \\
\hline
\end{tabular}
\caption{Summary of measurements of end-points in integrated invariant mass
distributions for the SPS1a benchmark SUSY model.
states. Column 2 lists the end-point positions expected from the masses
used in the {\tt HERWIG} generator while Columns 3 and 4 provide the
fitted end-point positions. Column 5 provides the systematic
uncertainty obtained from jet and lepton energy scale uncertainties of
1\% and 0.1\% respectively. All masses are in GeV. \label{tab1}}}

In order to determine the precision with which individual sparticle
masses can be measured with the global end-point constraints
listed above, 10000 LHC experiments were simulated with a simple toy
Monte Carlo code. For each MC experiment values for each of the
end-point positions were drawn from gaussian distributions centred on
the nominal values with widths determined by the uncorrelated
statistical and correlated systematic scale uncertainties listed in
Table~\ref{tab1}. The end-point positions were then fitted with the
formulae listed in Section~\ref{sec2.2}
(Eqns.~(\ref{eqn3}),~(\ref{eqn6}),~(\ref{eqn8}) and~(\ref{eqn10}))
using {\tt MINUIT} \cite{James:1975dr}. The $\chi^2$ minimisation
function which was used was that described in
Refs.~\cite{Allanach:2000kt,Gjelsten:2004ki}:
\begin{equation}
\chi^2=[{\bf E}^{\rm exp}-{\bf E}^{\rm th}({\bf m})]^T {\bf W}[{\bf
E}^{\rm exp}-{\bf E}^{\rm th}({\bf m})],
\end{equation}
where ${\bf E}^{\rm exp}$ and ${\bf E}^{\rm th}({\bf m})$ are vectors
of respectively observed and predicted end-point positions (the latter
functions of the vector of sparticle masses ${\bf m}$), and the weight
matrix ${\bf W}$ is the inverse of the correlation matrix of the
end-point observables. This form of the minimisation function takes
into account, through the use of ${\bf W}$, the correlations between
$\mqll$ and $\mql$ end-point observables generated by the correlated
JES systematic uncertainties.
\FIGURE[ht]{
\epsfig{file=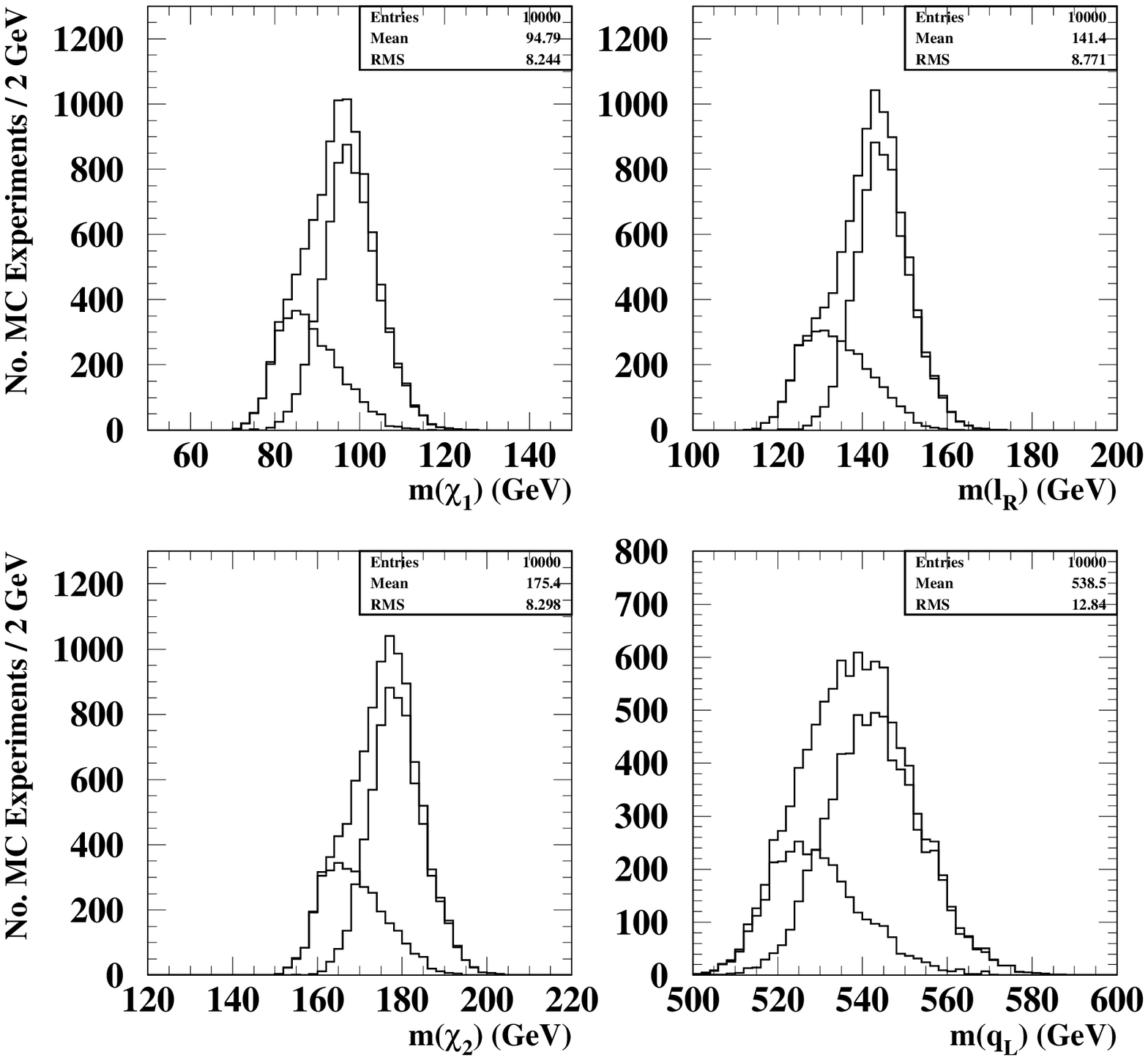,height=5.0in}
\caption{\label{fig8} Fitted sparticle masses for the SPS1a benchmark
SUSY model obtained from the conventional global end-point
analysis. In each case the two smaller histograms underlying the large
(sum) histogram represent the distributions obtained by using the
analytical mass formulae for cases A2 (left - incorrect assumption for
SPS1a) and A1 (right - correct assumption for SPS1a). The latter
distributions possess widths ranging from 6.6 GeV ($\mchioi$) to 11.3 GeV
($\msql$).} }

As discussed in Ref.~\cite{Gjelsten:2004ki}, excluding the $\mqllmin$
constraint generates a two-fold ambiguity in the allowed set of mass
values for the SPS1a model, corresponding to cases A1 and A2 in
Eqn.~(\ref{eqn6}). To properly account for this ambiguity we performed
the fit twice for each MC experiment, with the initial values of the
four masses set to the analytical solutions given by these two cases,
listed in Ref.~\cite{Gjelsten:2004ki}. The solution with the minimum
$\chi^2$ value was then selected.

The distributions of fitted mass values obtained from the 10000 toy MC
experiments are shown in Figure~\ref{fig8}, which also provides the
means and RMS values of the distributions. The latter provide
estimates of the sparticle mass measurement precisions. Also shown in
the Figure are the distributions of mass solutions obtained by
starting the fits from each of the two sets of analytical formulae
corresponding to cases A1 and A2. The incorrect solution corresponding
to case A2 contributes 32\% of solutions and generates mass values
$\sim$ 10 GeV below the nominal values. It can clearly be seen that
these incorrect solutions significantly decrease the precision with
which the masses can be measured.  

\subsection{Conditional end-point analysis}
\label{sec6.4}

When exploiting invariant mass correlations with
conditional end-points we start by confirming that the selected events
do indeed contain sequential two-body lepton-producing decays. In the
global end-point analysis this can be accomplished by studying the
shape of the $\mll$ distribution (Figure~\ref{fig7}(top)), which
departs from the canonical `smeared triangle' when the dilepton pair
is produced off mass-shell (see e.g. Ref.~\cite{Lester:2006cf}). This
shape analysis potentially requires large event statistics and a good
understanding of detector performance to conclusively exclude the
three-body hypothesis however, since the latter is effectively the off
mass-shell limit of the sequential two-body case. With invariant mass
correlations however we can also plot the reconstructed ratio
$\mqlhi/\mqllo$ for small $\mll$ discussed in
Section~\ref{sec4.3}. This is plotted in Figure~\ref{fig9} at
detector-level for $\mll$ $<$ 10 GeV. Comparison with
Figure~\ref{fig6} shows that the prominent peak associated with
sequential two-body lepton-producing decays has survived
detector-level smearing and the application of selection cuts and
hence this distribution provides clear evidence for such decay chains.
\FIGURE[ht]{
\epsfig{file=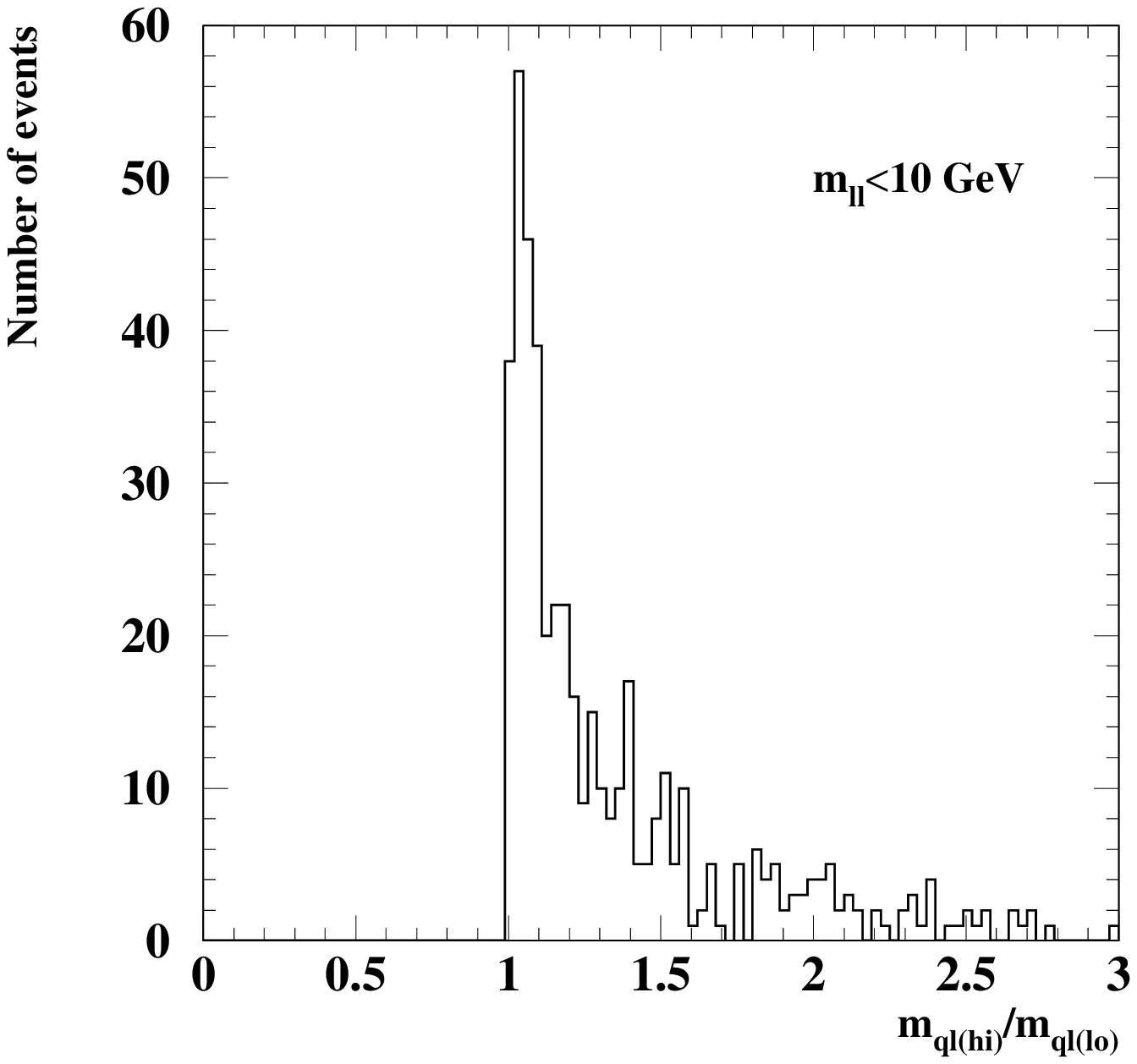,height=3.5in}
\caption{\label{fig9} Distribution of the detector-level ratio
$\mqlhi/\mqllo$ for $\mll$ $<$ 10 GeV for the SPS1a benchmark SUSY
model. } }  

The next step in the analysis involves identifying whether $\mqlhimax$
measures $\mqlfmax$ or $\mqlnmax$, i.e. whether cases A1/A2 or A3 in
Eqn.~(\ref{eqn6}) are correct. To accomplish this we construct two
$\mqlhi$ distributions (Figure~\ref{fig10}) -- one for events with low
$\mll$ ($<$ 30 GeV) and one for the remaining events (30 GeV $<$
$\mll$ $<$ $\mllmax$). Fitting the end-points $\mqlhimaxmll$ of these
distributions with simple linear functions we find that their
positions differ significantly (see Table~\ref{tab2}). This confirms
that the value of $\mqlhimax$ used in the global end-point
analysis measures $\mqlfmax$, as does the larger of the two
$\mqlhimaxmll$ end-points measured here. The smaller of the two
$\mqlhimaxmll$ end-points can then be used with Eqn.~(\ref{eqn29a}) to
determine $\mqlfmaxmllzero$ as described in Section~\ref{sec5}.
\FIGURE[ht]{
\epsfig{file=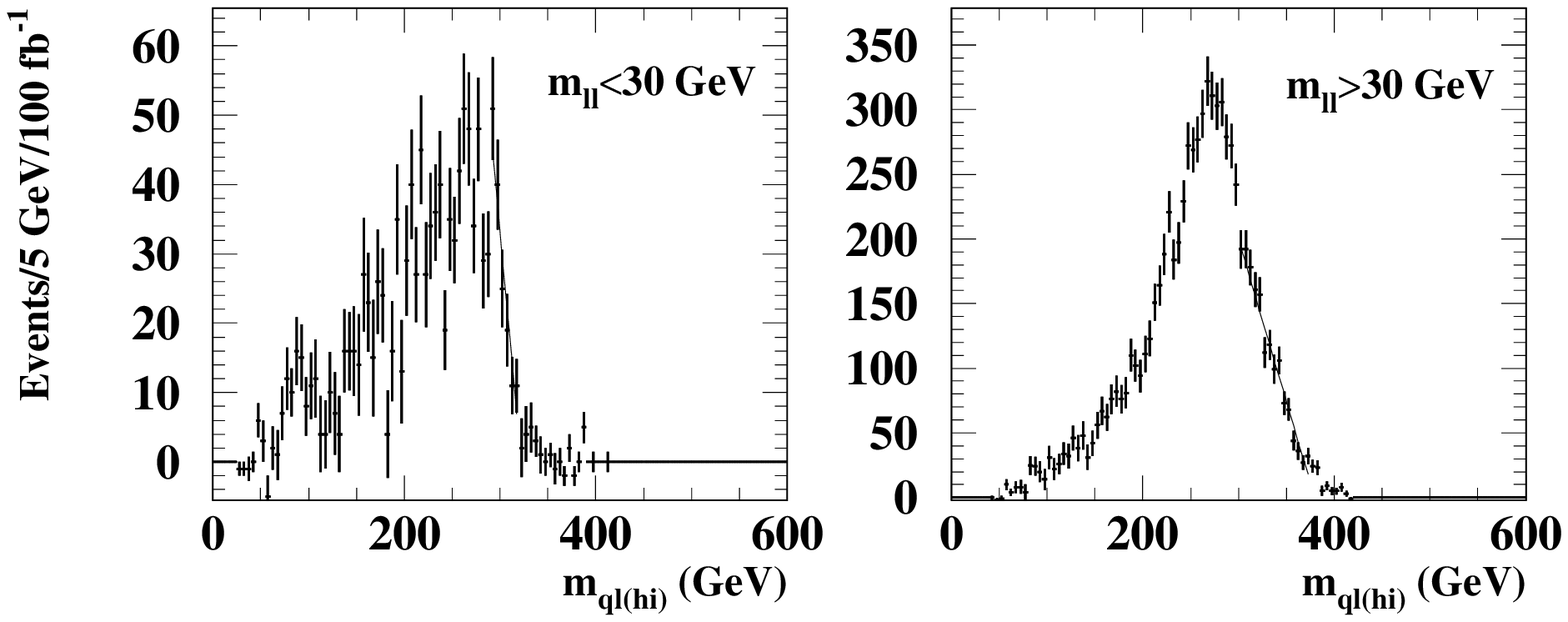,height=2.75in}
\caption{\label{fig10} Detector-level $\mqlhi$ distributions for the
SPS1a benchmark SUSY model for $\mll$ $<$ 30 GeV (left) and $\mll$ $>$
30 GeV (right). } }
\FIGURE[ht]{
\epsfig{file=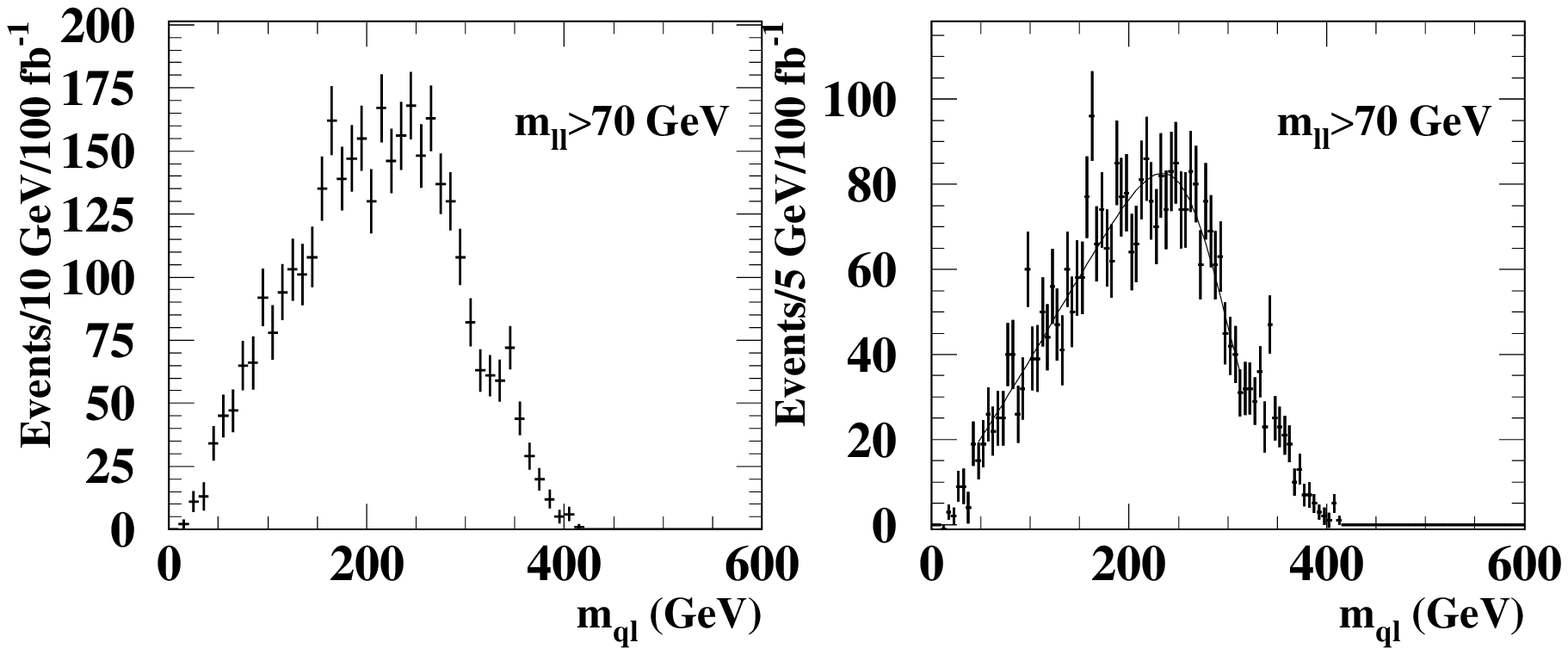,height=2.75in}
\caption{\label{fig11} Detector-level $\mql$ distribution for the
SPS1a benchmark SUSY model for $\mll$ $>$ 70 GeV (left) and showing
end-point fit (right). The same distribution is plotted in the
left-hand figure as the right-hand figure but with twice the bin-width
to improve the visibility of the $\mqln$ end-point.} }
\TABLE[ht]{\small%
\begin{tabular}{|c|c|c|c|c|c|c|}
\hline
Observable &$\mll$(low) &$\mll$(high) &Input &Mean &Statistical error &Energy scale error\\
\hline
$\mllmax$  & -- & --              &77.053  &77.006  &0.057 &0.077 \\ 
$\mqllmaxmll$ &0.0 &70.0          &428.4   &434.9   &1.4    &4.3 \\
$\mqllmaxmll$ &70.0 &$\mllmax$  &401.6   &400.0  &3.7    &4.0 \\
$\mqllmin$ &$\mllmax/\sqrt{2}$ &$\mllmax$&201.8   &200.8   &3.1    &2.0 \\
$\mqlhimaxmll$&0.0 &30.0          &317.6   &322.1   &3.2    &3.2 \\
$\mqlhimaxmll$&30.0 &$\mllmax$  &377.9   &379.6   &1.8    &3.8 \\
$\mqlnmax$ &70.0 &$\mllmax$  &300.2   &295.4   &2.7    &3.0 \\
$\mqllomaxmll$&0.0 &70.0          &300.2   &300.5   &1.2    &3.0 \\
$\mqllomaxmll$&70.0 &$\mllmax$  &268.3   &268.1   &2.6    &2.7 \\
\hline
\end{tabular}
\caption{Summary of measurements of end-points in conditional
invariant mass distributions for the SPS1a benchmark SUSY
model. Columns 2 and 3 provide the $\mll$ ranges over which the
invariant mass distributions are integrated to give the conditional
end-points listed in Column 1. Column 4 lists the end-point positions
expected from the masses used in the {\tt HERWIG} generator while
Columns 5 and 6 provide the fitted end-point positions and statistical
uncertainaties. Column 7 provides the systematic uncertainty obtained
from jet and lepton energy scale uncertainties of 1\% and 0.1\%
respectively. All masses are in GeV. \label{tab2}}}

Having excluded case A3 from Eqn.~(\ref{eqn6}) we now aim to measure
$\mqlnmax$ unambiguously, which will help us to resolve the A1/A2
ambiguity in $\mqllomax$. To do this we construct the distribution of
$\mql$, where each selected event contributes two entries ($\mqlhi$
and $\mqllo$). We further require that $\mll$ be large (70 GeV $<$
$\mll$ $<$ $\mllmax$) in order to maximise the separation between the
conditional upper bound on $\mqlhi$ provided by $\mqlfmaxmll$, at the
lower end of the $\mll$ range, and the `hidden' end-point measuring
$\mqlnmax$, whose position is independent of $\mll$. The value of
$\mqlfmaxmll$ at the lower end of the $\mll$ range determines the
value of $\mql$ at which the distribution begins to fall off $\sim$
linearly towards $\mqlfmax$. Consequently by maximising the separation
of this $\mqlfmaxmll$ value from the $\mqlnmax$ end-point we maximise
the visibility of the latter. The resulting $\mql$ distribution is
shown in Figure~\ref{fig11} and displays both the linear end-point
expected from $\mqlfmaxmll$ at 340 -- 380 GeV and the hidden
$\mqlnmax$ end-point at $\sim$ 300 GeV. The lack of dependence of the
$\mqlnmax$ end-point on $\mll$ causes it to have the same
characteristic `triangular' shape as the $\mllmax$ end-point at
parton-level and consequently we fit this distribution with the same
`smeared triangle' used to determine the $\mllmax$ end-point
position. The fit results are shown in Table~\ref{tab2}. The
visibility of the hidden $\mqlnmax$ end-point may be less clear if a
more realistic detector simulation is used, however more detailed
studies of the $\mql$ distribution over the full $\mll$ range may
corroborate an ambiguous observation in this case. Henceforth we shall
therefore assume that an unambiguous measurement can be obtained.

With the extra information provided by the $\mqlnmax$ end-point we
should potentially be able to resolve the A1/A2 ambiguity in
$\mqllomax$. Rather than simply measuring the value of the global
$\mqllo$ bound however we proceed by measuring the value
of the conditional bound $\mqllomaxmll$ in two $\mll$ bins: 0 $<$
$\mll$ $<$ 70 GeV and 70 GeV $<$ $\mll$ $<$ $\mllmax$. The first bin
contains the global end-point $\mqllomax$ (which for SPS1a measures
$\mqlnmax$ via case A1) while the second provides a measure of the
shape of the upper bound on $\mqllo$ determined by
Eqn.~(\ref{eqn100}). The $\mqllo$ distributions in these bins are
shown in Figure~\ref{fig12} and the results of linear end-point fits
are listed in Table~\ref{tab2}. Of course we do not know {\it a
priori} how to interpret these end-points, however by using these
measurements as constraints in the subsequent global mass fit the
correct interpretation can be determined with the help of the other
end-point measurements, particularly $\mqlnmax$. 
\FIGURE[ht]{
\epsfig{file=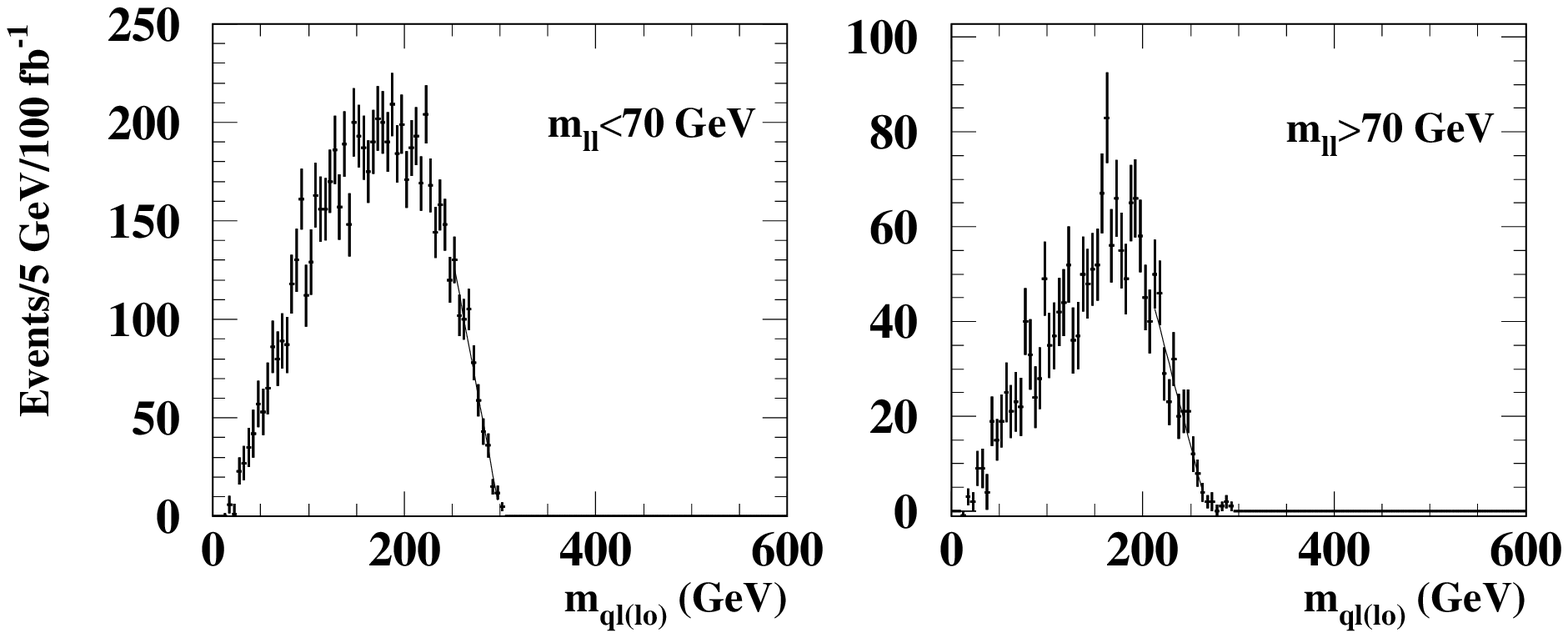,height=2.75in}
\caption{\label{fig12} Detector-level $\mqllo$ distributions for the
SPS1a benchmark SUSY model for $\mll$ $<$ 70 GeV (left) and $\mll$ $>$
70 GeV (right). } } 
\FIGURE[ht]{
\epsfig{file=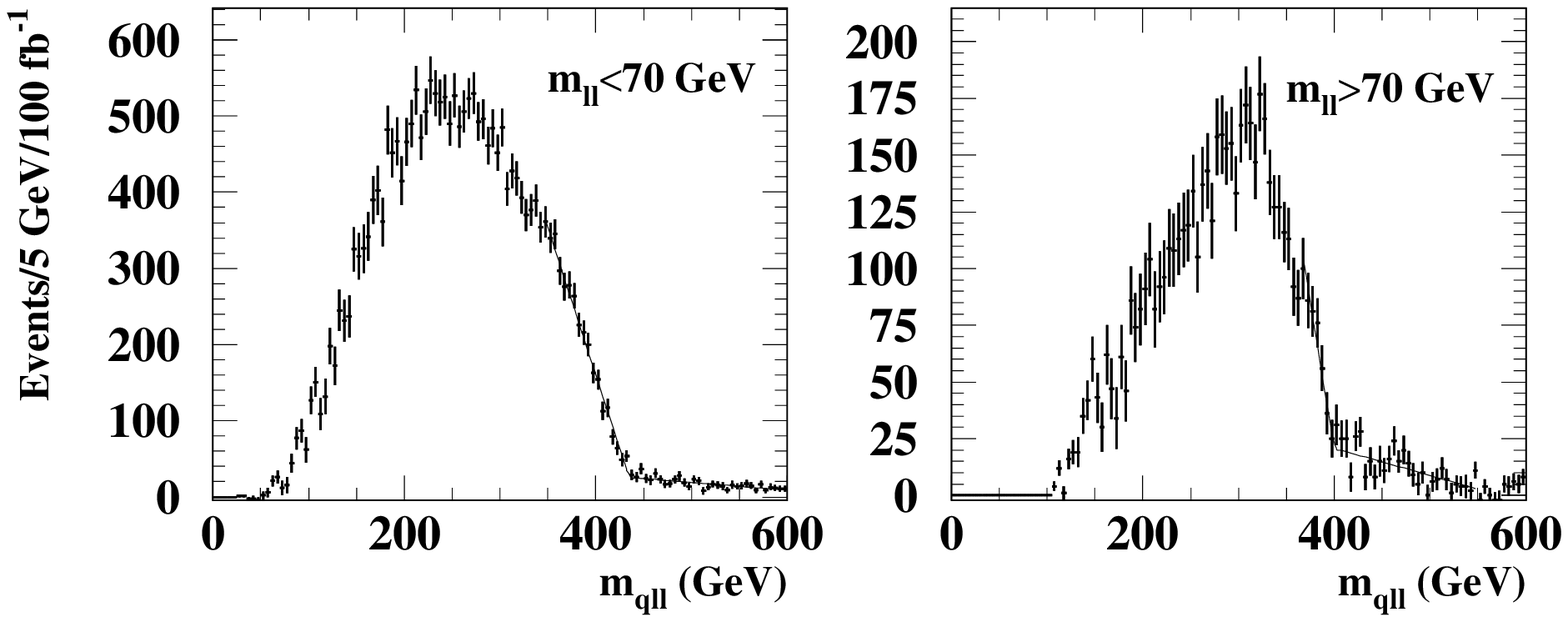,height=2.75in}
\caption{\label{fig13} Detector-level $\mqll$ distributions for the
SPS1a benchmark SUSY model for $\mll$ $<$ 70 GeV (left) and $\mll$ $>$
70 GeV (right). } }

We also measure the conditional bound $\mqllmaxmll$ in the same two
$\mll$ bins used above to measure $\mqllomaxmll$. The two $\mqll$
distributions are shown in Figure~\ref{fig13} while the results of
linear end-point fits are listed in Table~\ref{tab2}. The
$\mqllmaxmll$ end-point observed in the first bin lies at the position
of the global $\mqll$ end-point $\mqllmax$ while the end-point in the
second bin provides a measure of $\mqllmaxmll$ determined by
Eqn.~(\ref{eqn32}).

In addition to the above end-point constraints we make use of the
$\mllmax$ and $\mqllmin$ constraints from the global end-point
analysis. When added to the end-point constraints discussed above this
gives nine end-point constraints, listed in Table~\ref{tab2}. In most
cases there is reasonable agreement between the fitted means and the
nominal end-point positions. The fitted position of the first
$\mqllmaxmll$ end-point departs from the nominal position by 4.6$\sigma$
indicating that more work is needed to define the fitting function in
this case, however this will not be considered further here.

As in the global end-point analysis the precision with which SUSY
particle masses can be measured was determined by generating 10000 toy
MC experiments in which the end-point positions were smeared from their
nominal values with gaussians with widths determined by the
appropriate uncorrelated statistical and correlated systematic scale
uncertainties. A global fit to the end-point constraints was then
performed with the formulae from Sections~\ref{sec2.2} and~\ref{sec4.2} as
described in Section~\ref{sec6.3}. The initial values of the sparticle
masses used in the fits were determined by calculating $\mqlfmax$
(given by the larger $\mqlhimaxmll$ end-point), $\mqlfmaxmllzero$
(obtained from the two $\mqlhimaxmll$ end-points), $\mqlnmax$ and
$\mllmax$ and then using Eqns.~(\ref{eqn101})--(\ref{eqn104}).

The distributions of sparticle masses obtained from the 10000 MC
experiments are listed in Figure~\ref{fig14}. The means and RMS values
of these distributions are compared with the equivalent quantities
from the global end-point analysis in Table~\ref{tab3}. The
distributions obtained from the conditional end-point analysis are
$\sim$ 20\%--30\% narrower, and also less skewed, than those obtained
using global end-points, primarily due to resolution of the A1/A2
ambiguity but also due to the use of additional end-point
constraints. It should be noted however that as in the case of the
global end-point analysis the conditional end-point analysis does not
establish the identities of the intermediate states, merely their
masses.
\FIGURE[ht]{
\epsfig{file=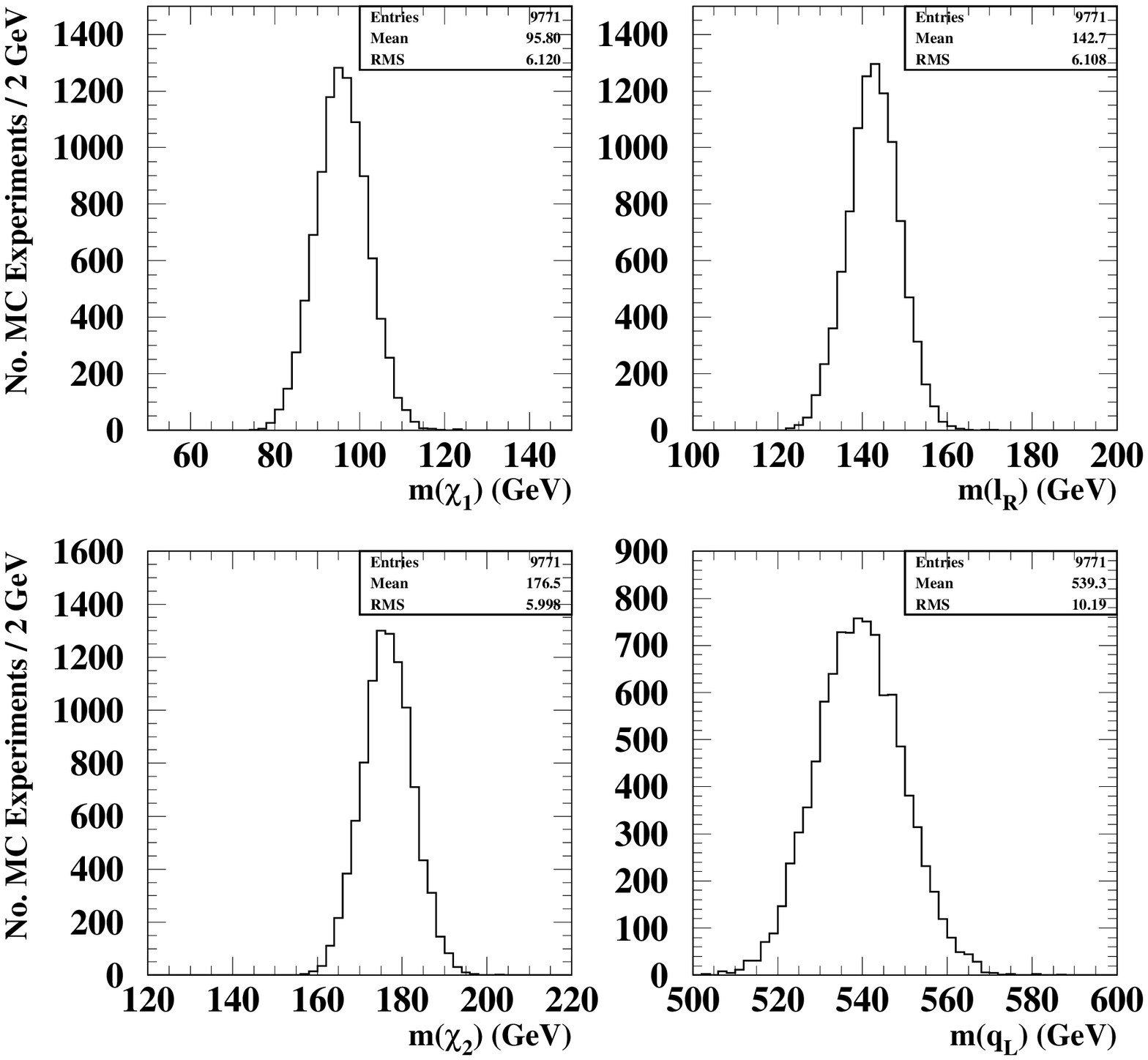,height=5.0in}
\caption{\label{fig14} Fitted sparticle masses for the SPS1a
benchmark SUSY model obtained from the conditional end-point analysis using invariant mass correlations. } }
\TABLE[ht]{\small%
\begin{tabular}{|c|c|c|c|c|c|}
\hline
State &Input &\multicolumn{2}{|c|}{Global end-point fit} &\multicolumn{2}{|c|}{Conditional end-point fit}\\
\cline{3-6}
 & &Mean &Error &Mean &Error \\
\hline
$\chioi$  &96.05       &94.8  &8.2  &95.8   &6.1 \\ 
$\slr$    &142.97      &141.4 &8.8  &142.7  &6.1 \\
$\chioii$ &176.81      &175.4 &8.3  &176.5  &6.0 \\
$\sql$    &540.0&538.5 &12.8 &539.3  &10.2 \\
\hline
\end{tabular}
\caption{Summary of sparticle mass measurement precisions for SPS1a
states. Column 2 lists input masses used in the toy MC simulation,
Columns 3 and 4 the fitted masses and uncertainties obtained from the
conventional global end-point analysis and Columns 5 and 6 the
equivalent values obtained from the conditional end-point fit. All masses are in GeV.  \label{tab3}}}

\section{Conclusions}
\label{sec7}

This paper has shown that additional constraints on sparticle
masses at the LHC can be obtained by exploiting correlations between
invariant mass end-points obtained from the sequential two-body decay
chains used in the conventional global end-point analysis. The same
techniques can also be used to confirm that selected events contain
sequential two-body lepton-producing decays rather than three-body
decays. These additional constraints can be used to improve the precision
with which sparticle masses can be measured, although care must be
taken to acccount for any correlations between end-point positions. 

The results of this paper indicate a number of potential avenues for
future work. Most importantly a detailed detector-level Monte Carlo
study of a large number of experiments should be performed to
determine the level of correlation between different end-point
observables. With a more detailed understanding of such correlations
more constraints could be used than in the simple study
described here, leading to further improvements in mass measurement
precision. In addition it would be profitable to calculate analytical
formulae for the end-point shapes similar to those for the
global end-points described in Ref.~\cite{Miller:2005zp}.

\section*{Acknowledgements}
The authors wish to thank Giacomo Polesello for helpful comments on a
draft of this paper and for providing the simulated datasets on which
the work was based. They also wish to thank Chris Lester for further
helpful comments. DRT and DC wish to acknowledge STFC for support.

\renewcommand{\theequation}{A\arabic{equation}}
\setcounter{equation}{0}
\section*{Appendix A: Three-body lepton-producing decay chains}

For completeness we list here some formulae for kinematic bounds
equivalent to those listed in Section~\ref{sec4}, for three-body
lepton-producing decay chains incorporating single two-body
quark-producing decays of the form:
\begin{equation}
\label{eqnA1}
  \sql \rightarrow \chioii q \rightarrow \chioi l_1l_2q.
\end{equation}
The global end-points from such decay chains in NUHM models were
considered in Ref.~\cite{Lester:2006cf}.

The $\mqll$ bound as a function of $\mll$ in this case is identical
to that for the two-body case listed in Eqn.~(\ref{eqn32}). The
$\mqlhi$ bound as a function of $\mll$ equivalent to
Eqns.~(\ref{eqn29a}) and~(\ref{eqn29b}) can also be obtained from
Eqn.~(\ref{eqn32}) by recognising that $\mqlhi$ is maximised when the
$\chioi$, $q$ and one of the leptons are all co-linear in the rest
frame of the $\chioii$. In this case one of the two lepton+quark
invariant masses is zero and hence
\begin{eqnarray}
\label{eqnA2}
\big(\mqlhi^{{\rm bound}(\mll)}\big)^2&=&\big(\mqll^{{\rm bound}(\mll)}\big)^2-\mll^2 \\ 
&=& \frac{\big(\msql^2-\mchioii^2\big)}{2\mchioii^2}\Big(\mchioii^2-\mchioi^2+\mll^2 \pm \sqrt{\lambda(\mchioii,\mchioi,\mll)}\Big).
\end{eqnarray}
This is maximised when $\mll=0$, which gives the position of the
global end-point reported in Ref.~\cite{Lester:2006cf}:
\begin{eqnarray}
\label{eqnA3}
\big(\mqlhimax\big)^2 = \frac{\big(\msql^2-\mchioii^2\big)\big(\mchioii^2-\mchioi^2\big)}{\mchioii^2},
\end{eqnarray}
while if $\mll=\mllmax=\mchioii-\mchioi$ then 
\begin{eqnarray}
\label{eqnA4}
\big(\mqlhimaxmll\big)^2 = \frac{\big(\msql^2-\mchioii^2\big)\big(\mchioii^2-\mchioii\mchioi\big)}{\mchioii^2}.
\end{eqnarray}

The $\mqllo$ bound equivalent to that derived from Eqn.~(\ref{eqn100})
can be obtained from the three-body analogue of Eqn.~(\ref{eqn26})
by conserving energy and momentum in the $qll$ plane. This equation
is more complicated than Eqn.~(\ref{eqn26}) but setting
$m_{ql_1}=m_{ql_2}$ it simplifies considerably to give:
\begin{eqnarray}
\label{eqnA5}
\big(\mqllo^{{\rm bound}(\mll)}\big)^2&=& \frac{\big(\msql^2-\mchioii^2\big)}{4\mchioii^2}\Big(\mchioii^2-\mchioi^2+\mll^2 \pm \sqrt{\lambda(\mchioii,\mchioi,\mll)}\Big)\\
&=& \frac{\big(\mqlhi^{{\rm bound}(\mll)}\big)^2}{2}.
\end{eqnarray}
Consequently the $\mqlhi$ bound is $\sqrt{2}$ times as large as the
$\mqllo$ bound, as was noted for the global maximum in
Ref.~\cite{Lester:2006cf}.

\end{document}